\documentclass[apj]{emulateapj}
\usepackage{longtable}

\def\kms{\ifmmode{\rm km\thinspace s^{-1}}\else km\thinspace s$^{-1}$\fi}

\shortauthors{Torres et al.}
\shorttitle{Improved spectroscopic parameters}

\begin{document}


\title{Improved spectroscopic parameters for transiting planet hosts}

\author{
Guillermo Torres\altaffilmark{1},
Debra A.\ Fischer\altaffilmark{2},
Alessandro Sozzetti\altaffilmark{3},
Lars A.\ Buchhave\altaffilmark{4},
Joshua N.\ Winn\altaffilmark{5},
Matthew J.\ Holman\altaffilmark{1}, and
Joshua A.\ Carter\altaffilmark{1}
}

\altaffiltext{1}{Harvard-Smithsonian Center for Astrophysics, 60
Garden St., Cambridge, MA 02138, USA; e-mail: gtorres@cfa.harvard.edu}

\altaffiltext{2}{Dept.\ of Astronomy, Yale University, New Haven, CT
06511, USA}

\altaffiltext{3}{INAF -- Osservatorio Astronomico di Torino, 10025
Pino Torinese, Italy}

\altaffiltext{4}{Niels Bohr Institute, Copenhagen University, Denmark}

\altaffiltext{5}{Dept.\ of Physics, and Kavli Institute for
Astrophysics and Space Research, Massachusetts Institute of
Technology, Cambridge, MA 02139, USA}

\begin{abstract} 

  We report homogeneous spectroscopic determinations of the effective
  temperature, metallicity, and projected rotational velocity for the
  host stars of 56 transiting planets. Our analysis is based primarily
  on the Stellar Parameter Classification (SPC) technique. We
  investigate systematic errors by examining subsets of the data with
  two other methods that have often been used in previous studies (SME
  and MOOG). The SPC and SME results, both based on comparisons
  between synthetic spectra and actual spectra, show strong
  correlations between $T_{\rm eff}$, [Fe/H], and $\log g$ when
  solving for all three quantities simultaneously. In contrast the
  MOOG results, based on a more traditional curve-of-growth approach,
  show no such correlations. To combat the correlations and improve
  the accuracy of the temperatures and metallicities, we repeat the
  SPC analysis with a constraint on $\log g$ based on the mean stellar
  density that can be derived from the analysis of the transit light
  curves.  Previous studies that have not taken advantage of this
  constraint have been subject to systematic errors in the stellar
  masses and radii of up to 20\% and 10\%, respectively, which can be
  larger than other observational uncertainties, and which also cause
  systematic errors in the planetary mass and radius.

\end{abstract}

\keywords{
planetary systems ---
stars: abundances ---
stars: fundamental parameters ---
techniques: spectroscopic
}

\section{Introduction}
\label{sec:introduction}

In recent years the number of extrasolar transiting planets has
expanded considerably, with discoveries being made at a rapid pace
both from the ground and increasingly also from space by the
\emph{CoRoT} and \emph{Kepler} missions. With this large assembly of
data, studies have begun to focus on examining patterns and
correlations among the global properties of these planets and their
parent stars, and what this can tell us about planet formation and
evolution.

While the characteristics of some of these systems are very well known
(e.g., HD\,209458, HD\,189733, TrES-1), those of others are much less
well determined and have remained so since their discovery.  Our
knowledge of the planetary properties depends critically on
understanding the parent stars. This is because, for transiting
systems, the light curves only give information on the size of the
planet relative to that of the star ($R_p \propto R_{\star}$), and
spectroscopic observations only reveal the mass of the planet if we
know the mass of the star ($M_p \propto M_{\star}^{2/3}$). The stellar
mass and radius, in turn, depend on other properties that can be
gleaned from the stars' spectra such as the effective temperature
($T_{\rm eff}$), surface gravity ($\log g$), and chemical composition
(commonly represented by [Fe/H]).

For many of the known transiting planet systems, follow-up photometric
observations have been undertaken after the initial discovery of the
planet, usually for the purpose of measuring the times of mid-transit
and seeking departures from strict periodicity (transit timing
variations) that may indicate the presence of additional bodies in the
system. These new transit light curve observations have also served to
improve the radius determinations in many cases. However, it is much
less common for known transiting systems to be re-observed
spectroscopically. As a result, our knowledge of the stellar
properties is often limited by whatever information was reported in
the discovery papers, which is sometimes preliminary. Inaccuracies in
the stellar $T_{\rm eff}$, $\log g$, and [Fe/H] propagate through to
the determination of the planetary properties, and may obscure
correlations with other quantities and prevent us from gaining
valuable insight into the nature of planets. To make matters worse,
the methods of determining $T_{\rm eff}$, $\log g$, and [Fe/H] in the
literature are highly inhomogeneous, as they have been carried out by
many groups using different assumptions and methodologies.

In one of the few studies to redetermine spectroscopic properties for
transiting planet hosts in a uniform way, \cite{Ammler:09} obtained
new temperatures, surface gravities, and metallicities for 13 systems
based on new or existing spectra. They combined their determinations
with those for 11 additional systems made by others using similar
techniques, and compiled a list of 24 host stars with uniformly
derived properties. Comparison with results for the same stars by
other groups uncovered significant systematic differences in some
cases, the causes of which are unknown.

One of the motivations for the present paper is to derive
spectroscopic properties in a homogeneous manner for a much larger
sample of more than 50 transiting planet hosts, and thereby reduce any
dispersion in the stellar and planetary properties that is caused by
the variety of methodologies used in the past. For this we make use of
the Stellar Parameter Classification (SPC) technique introduced by
\cite{Buchhave:12}. We also wish to understand potential systematic
errors in such determinations. In particular, it has been recognized
for some time \citep[e.g.,][]{Sozzetti:07, Holman:07, Winn:08} that
surface gravities are often poorly constrained by spectral analyses.
This is an unfortunate limitation, because the surface gravity would
otherwise help to establish the luminosity of the star (and hence its
radius) by placing it on the H-R diagram.

In cases for which the surface gravity is poorly constrained, the use
of an external constraint on the luminosity becomes very important to
allow accurate determinations of the mass and radius of the
star. However, a detail that has usually been overlooked is that the
determinations of other spectroscopic quantities such as the
temperature and metallicity can \emph{also} be affected by the poor
constraint on gravity. This is because the uncertainties in $T_{\rm
eff}$ and [Fe/H] are strongly correlated with $\log g$ in at least
some of the commonly used analysis techniques.  This can be a
significant source of systematic error. Therefore, a second goal of
our work is to compare spectroscopic determinations for a subset of
the sample obtained using two additional procedures widely employed in
previous studies, and to investigate and quantify systematics errors
stemming from the weakly constrained gravities.

Ultimately the stellar quantities that enter into the calculation of
the planetary characteristics are the masses and radii inferred from
$T_{\rm eff}$, $\log g$, and [Fe/H], as mentioned earlier. A third
objective of the present paper is to study and quantify how errors in
the spectroscopic properties propagate into the stellar masses and
radii.

The paper is organized as follows. In Sect.~\ref{sec:observations} we
report our spectroscopic observations, which consist of new and
archival spectra obtained with three different telescopes. Our
spectroscopic analysis techniques are described in
Sect.~\ref{sec:analysis}. Our results with and without the application
of external constraints on the surface gravity are presented in
Sect.~\ref{sec:results}. Also presented there are the final results
from this work. These are then compared with work by others in
Sect.~\ref{sec:comparison}. In Sect.~\ref{sec:impact} we study the
impact of the different assumptions regarding $\log g$ on the stellar
masses and radii. We discuss our results in
Sect.~\ref{sec:discussion}, and end with a summary of our conclusions
in Sect.~\ref{sec:conclusions}.

\section{Spectroscopic observations and reductions}
\label{sec:observations}

For this work we have relied in part on new spectra collected with the
TRES instrument \citep{Furesz:08} on the 1.5\,m Tillinghast reflector
at the F.\ L.\ Whipple Observatory (Mount Hopkins, AZ), and also on
publicly available spectra obtained with the HIRES instrument
\citep{Vogt:94} on the Keck~I telescope (Mauna Kea, HI) and with the
FIES instrument \citep{Djupvic:10} on the 2.5\,m Nordic Optical
Telescope (La Palma, Spain). The archival spectra have generally not
been used previously for a determination of the stellar properties,
and even when they have, we have redetermined those properties here
using different methodologies for comparison purposes.

We obtained new TRES spectra for 22 stars (see
Table~\ref{tab:tresstars}).  They cover the wavelength range
$\sim$3860--9100\,\AA\ and were taken at a typical resolving power of
$R \approx 48,\!000$ with the medium fiber of that instrument. The
signal-to-noise ratios (SNRs) for individual exposures range between
27 and 130 per resolution element of 6.2\,\kms, and refer to the
region of the \ion{Mg}{1}\,b triplet ($\sim$5200\,\AA).  For stars
with multiple exposures the SNR reported is the average.  All spectra
were reduced using the procedures described by \cite{Buchhave:10a,
Buchhave:10b}.

\begin{deluxetable}{lccc@{}ccc@{}ccc@{}c}
\tablewidth{19pc}
\tablecaption{Exoplanet host stars with new TRES spectra.
\label{tab:tresstars}}
\tablehead{
\colhead{} &
\colhead{} &
\colhead{} &
\multicolumn{2}{c}{SPC} &&
\multicolumn{2}{c}{SME} &&
\multicolumn{2}{c}{MOOG} \\
\colhead{} &
\colhead{} &
\colhead{} &
\multicolumn{2}{c}{analysis} &&
\multicolumn{2}{c}{analysis} &&
\multicolumn{2}{c}{analysis} \\ [+0.5ex]
\cline{4-5} \cline{7-8} \cline{10-11} \\ [-1.5ex]
\colhead{Star} &
\colhead{$N_{\rm obs}$} &
\colhead{SNR} &
\colhead{U} & \colhead{C} &&
\colhead{U} & \colhead{C} &&
\colhead{U} & \colhead{C}
}
\startdata
 HAT-P-3  & 3 & 36  &   X & X   &&  -- & -- &&   -- & -- \\
 HAT-P-4  & 1 & 51  &   X & X   &&  -- & -- &&   -- & -- \\
 HAT-P-5  & 1 & 44  &   X & X   &&  -- & -- &&    X & X \\
 HAT-P-7  & 1 & 68  &   X & X   &&  -- & -- &&   -- & -- \\
 HAT-P-9  & 4 & 42  &   X & X   &&  -- & -- &&    X & X \\
 HAT-P-10 & 1 & 27  &   X & X   &&  -- & -- &&   -- & -- \\
 HAT-P-13 & 1 & 38  &   X & X   &&  -- & -- &&   -- & -- \\
 HAT-P-15 & 1 & 42  &   X & X   &&  -- & -- &&   -- & -- \\
 HAT-P-17 & 1 & 56  &   X & X   &&  -- & -- &&   -- & -- \\
 HAT-P-19 & 1 & 37  &   X & X   &&  -- & -- &&   -- & -- \\
 HAT-P-20 & 1 & 30  &   X & X   &&  -- & -- &&   -- & -- \\
 HAT-P-21 & 1 & 37  &   X & X   &&  -- & -- &&   -- & -- \\
 HAT-P-22 & 1 & 81  &   X & X   &&  -- & -- &&   -- & -- \\
 HAT-P-24 & 1 & 57  &   X & X   &&  -- & -- &&   -- & -- \\
 HAT-P-26 & 1 & 51  &   X & X   &&  -- & -- &&   -- & -- \\
 HAT-P-29 & 1 & 48  &   X & X   &&  -- & -- &&   -- & -- \\
HD 147506 & 2 & 98  &   X & X   &&  -- & -- &&   -- & -- \\   
 WASP-2   & 1 & 44  &   X & X   &&  -- & -- &&    X & X \\
 WASP-3   & 1 & 62  &   X & X   &&  -- & -- &&    X & X \\
 WASP-10  & 2 & 35  &   X & X   &&  -- & -- &&    X & X \\
 WASP-13  & 3 & 78  &   X & X   &&  -- & -- &&    X & X \\
 WASP-14  & 4 & 104 &   X & X   &&  -- & -- &&    X & X \\ [-2ex]
\enddata

\tablecomments{U = unconstrained analysis ($\log g$ free); C =
constrained analysis ($\log g$ fixed to best value from photometry); X
indicates the method has been applied to this star.}

\end{deluxetable}

The archival HIRES spectra (Table~\ref{tab:hiresstars}) were obtained
with the typical setup employed for extrasolar planet searches with
that instrument.  These spectra cover the full optical range and were
gathered at two different nominal resolving powers of $R \approx
68,\!000$
and $R \approx 51,\!000$.
The SNRs in the Mg\,b region range from about 40 to 360 per resolution
element (4.4\,\kms\ and 5.9\,\kms, respectively). Many of the spectra
available from this instrument used an iodine cell in the beam to
impose a dense set of molecular absorption lines that serve as a
fiducial in determining precise Doppler shifts.  For the present work
we analyzed only the spectra taken without the iodine cell (referred
to as ``templates''). They were reduced following standard procedures
for this instrument.

\begin{deluxetable}{lccc@{}ccc@{}ccc@{}c}
\tablewidth{19pc}
\tablecaption{Exoplanet host stars with archival HIRES spectra.
\label{tab:hiresstars}}
\tablehead{
\colhead{} &
\colhead{} &
\colhead{} &
\multicolumn{2}{c}{SPC} &&
\multicolumn{2}{c}{SME} &&
\multicolumn{2}{c}{MOOG} \\
\colhead{} &
\colhead{} &
\colhead{} &
\multicolumn{2}{c}{analysis} &&
\multicolumn{2}{c}{analysis} &&
\multicolumn{2}{c}{analysis} \\ [+0.5ex]
\cline{4-5} \cline{7-8} \cline{10-11} \\ [-1.5ex]
\colhead{Star} &
\colhead{$N_{\rm obs}$} &
\colhead{SNR} &
\colhead{U} & \colhead{C} &&
\colhead{U} & \colhead{C} &&
\colhead{U} & \colhead{C}
}
\startdata
 CoRoT-1    & 1 &  84  &    X & X   && X & X   &&   X & X \\
 CoRoT-2    & 1 & 128  &    X & X   && X & X   &&   X & X \\
 CoRoT-7    & 1 & 103  &   -- & X   && X & X   &&   X & X \\
 HAT-P-3    & 1 & 163  &    X & X   && L & --  &&  -- & -- \\
 HAT-P-4    & 1 & 202  &    X & X   && L & L   &&  -- & -- \\
 HAT-P-6    & 1 & 303  &    X & X   && X & X   &&   X & X \\
 HAT-P-7    & 1 & 344  &    X & X   && L & --  &&  -- & -- \\
 HAT-P-8    & 1 & 276  &    X & X   && L & L   &&  -- & -- \\
 HAT-P-10   & 1 & 179  &    X & X   && L & --  &&  -- & -- \\
 HAT-P-11   & 3 & 319  &    X & X   && L & L   &&  -- & -- \\
 HAT-P-13   & 3 & 192  &    X & X   && L & L   &&  -- & -- \\
 HAT-P-14   & 1 & 284  &    X & X   && L & L   &&  -- & -- \\
 HAT-P-15   & 1 & 148  &    X & X   && L & L   &&  -- & -- \\
 HAT-P-16   & 1 & 206  &    X & X   && L & L   &&  -- & -- \\
 HAT-P-17   & 1 & 247  &    X & X   && L & L   &&  -- & -- \\
 HAT-P-18   & 1 & 125  &    X & X   && L & L   &&  -- & -- \\
 HAT-P-19   & 1 & 104  &    X & X   && L & L   &&  -- & -- \\
 HAT-P-21   & 1 & 160  &    X & X   && L & L   &&  -- & -- \\
 HAT-P-22   & 1 & 298  &    X & X   && L & L   &&  -- & -- \\
 HAT-P-23   & 1 & 156  &    X & X   && L & L   &&  -- & -- \\
 HAT-P-24   & 2 & 187  &    X & X   && L & L   &&  -- & -- \\
 HAT-P-25   & 2 & 124  &    X & X   && L & L   &&  -- & -- \\
 HAT-P-26   & 1 & 165  &    X & X   && L & --  &&  -- & -- \\ 
 HD 17156   & 1 & 288  &    X & X   && X & X   &&   X & X \\
 HD 80606   & 1 & 357  &    X & X   && X & X   &&   X & X \\
 HD 147506  & 1 & 330  &    X & X   && X & X   &&   X & X \\
 HD 149026  & 4 & 360  &    X & X   && X & X   &&   X & X \\
 HD 189733  & 3 & 337  &    X & X   && X & X   &&   X & X \\
 Kepler-6   & 1 &  97  &    X & X   && X & X   &&   X & X \\
 Kepler-7   & 2 &  69  &    X & X   && X & X   &&   X & X \\
 Kepler-8   & 1 &  85  &    X & X   && X & X   &&   X & X \\
 Kepler-9   & 1 &  63  &    X & --  && X & --  &&   X & -- \\
 Kepler-10  & 1 & 226  &    X & X   && X & X   &&   X & X \\
 Kepler-11  & 1 &  41  &    X & --  && X & --  &&   X & -- \\
 TRES-1     & 2 & 109  &    X & X   && X & X   &&   L & X \\
 TRES-2     & 2 & 178  &    X & X   && X & X   &&   L & X \\
 TRES-3     & 1 &  97  &    X & X   && X & X   &&   L & X \\
 WASP-1     & 3 & 184  &    X & X   && X & X   &&   X & X \\
 WASP-2     & 1 & 161  &    X & X   && X & X   &&   X & X \\
 WASP-3     & 1 & 252  &    X & X   && X & X   &&   X & X \\
 WASP-12    & 1 & 136  &    X & X   && X & X   &&   X & X \\
 WASP-13    & 1 & 150  &    X & X   && X & X   &&   X & X \\
 WASP-14    & 1 & 265  &    X & X   && X & X   &&   X & X \\
 WASP-17    & 1 & 103  &    X & X   && X & X   &&   X & X \\
 WASP-18    & 1 & 153  &    X & X   && X & X   &&   X & X \\
 WASP-19    & 1 & 100  &    X & X   && X & X   &&   X & X \\
 XO-1       & 1 & 147  &    X & X   && X & X   &&   X & X \\
 XO-2       & 1 & 180  &    X & X   && X & X   &&   X & X \\
 XO-3       & 1 & 243  &    X & X   && X & X   &&   X & X \\
 XO-4       & 1 & 222  &    X & X   && X & X   &&   X & X \\ [-2ex]
\enddata

\tablecomments{U = unconstrained analysis ($\log g$ free); C =
constrained analysis ($\log g$ fixed to best value from photometry); X
indicates the method has been applied to this star. L = results taken
from the literature and obtained with the same technique.}

\end{deluxetable}

The archival FIES spectra cover the wavelength range from about
3600\,\AA\ to 7400\,\AA, and were obtained with the medium fiber of
that spectrograph at a typical resolving power of $R \approx
46,\!000$. The SNRs at $\sim$5200\,\AA\ for individual exposures range
from 21 to 220 per resolution element of 6.5\,\kms. The list of stars
is given in Table~\ref{tab:fiesstars}.  Reductions were carried out as
described by \cite{Buchhave:10a, Buchhave:10b} using nightly
flatfield, bias, and dark frames, as well as thorium-argon exposures
taken immediately before and after the science exposure. This
instrument does not use an iodine cell, and relies instead on its
intrinsic stability to enable the measurement of precise radial
velocities of exoplanet hosts. As a result, there are usually many
more spectra of each object available for spectroscopic analysis.

\begin{deluxetable}{lccc@{}ccc@{}ccc@{}c}
\tablewidth{19pc}
\tablecaption{Exoplanet host stars with archival FIES spectra.
\label{tab:fiesstars}}
\tablehead{
\colhead{} &
\colhead{} &
\colhead{} &
\multicolumn{2}{c}{SPC} &&
\multicolumn{2}{c}{SME} &&
\multicolumn{2}{c}{MOOG} \\
\colhead{} &
\colhead{} &
\colhead{} &
\multicolumn{2}{c}{analysis} &&
\multicolumn{2}{c}{analysis} &&
\multicolumn{2}{c}{analysis} \\ [+0.5ex]
\cline{4-5} \cline{7-8} \cline{10-11} \\ [-1.5ex]
\colhead{Star} &
\colhead{$N_{\rm obs}$} &
\colhead{SNR} &
\colhead{U} & \colhead{C} &&
\colhead{U} & \colhead{C} &&
\colhead{U} & \colhead{C}
}
\startdata
   HAT-P-6 & 34 &  87 &  X & X   && -- & -- &&   -- & -- \\
   HAT-P-8 &  2 & 121 &  X & X   && -- & -- &&   -- & -- \\
   HAT-P-9 &  8 &  32 &  X & X   && -- & -- &&   -- & -- \\
  HAT-P-10 &  1 &  24 &  X & X   && -- & -- &&   -- & -- \\
  HAT-P-13 & 35 &  92 &  X & X   && -- & -- &&   -- & -- \\
  HAT-P-19 &  2 &  37 &  X & X   && -- & -- &&   -- & -- \\
  HAT-P-25 &  1 &  21 &  X & X   && -- & -- &&   -- & -- \\
  HAT-P-29 &  2 &  36 &  X & X   && -- & -- &&   -- & -- \\
 HD 147506 &  1 & 204 &  X & X   && -- & -- &&   -- & -- \\  
    WASP-1 & 33 &  77 &  X & X   && -- & -- &&   -- & -- \\
    WASP-2 &  7 &  86 &  X & X   && -- & -- &&   -- & -- \\
    WASP-3 &  9 & 103 &  X & X   && -- & -- &&   -- & -- \\
   WASP-10 & 14 &  53 &  X & X   && -- & -- &&   -- & -- \\
   WASP-11 &  3 &  59 &  X & X   && -- & -- &&   -- & -- \\
   WASP-12 & 22 &  74 &  X & X   && -- & -- &&   -- & -- \\
   WASP-13 &  5 & 119 &  X & X   && -- & -- &&   -- & -- \\
   WASP-14 & 10 & 152 &  X & X   && -- & -- &&   -- & -- \\
   WASP-24 &  5 &  67 &  X & X   && -- & -- &&   -- & -- \\
   WASP-31 & 23 &  35 &  X & X   && -- & -- &&   -- & -- \\
   WASP-38 & 27 & 203 &  X & X   && -- & -- &&   -- & -- \\
      XO-3 & 40 &  98 &  X & X   && -- & -- &&   -- & -- \\
      XO-4 & 28 &  59 &  X & X   && -- & -- &&   -- & -- \\  [-2ex]
\enddata

\tablecomments{U = unconstrained analysis ($\log g$ free); C =
constrained analysis ($\log g$ fixed to best value from photometry); X
indicates the method has been applied to this star.}

\end{deluxetable}

\begin{figure} 
\epsscale{1.2} 
\plotone{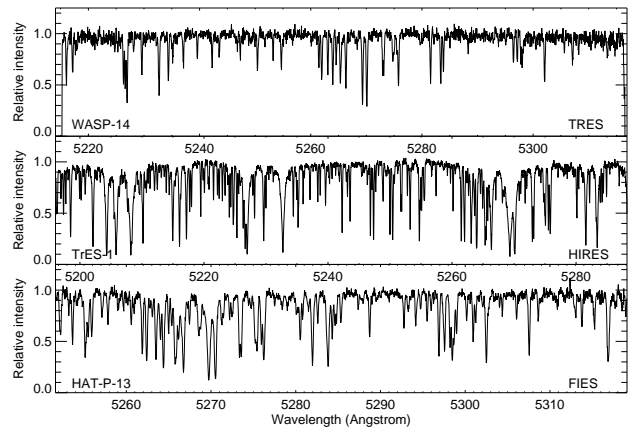}
\figcaption[]{Sample spectra near the \ion{Mg}{1}\,b order from each
of the three instruments used in this work, as labeled. The
signal-to-noise ratios per resolution element are 129 (WASP-14), 109
(TRES-1), and 43 (HAT-P-13).
\label{fig:spectra}}
\end{figure}

Sample spectra from each of the instruments are shown in
Figure~\ref{fig:spectra}.

\section{Spectroscopic analysis techniques}
\label{sec:analysis}

Three different techniques were used to determine the spectroscopic
parameters $T_{\rm eff}$, $\log g$, and [Fe/H]. One procedure was
applied to all of our spectra, to provide a homogeneous dataset.  The
other two techniques were applied to subsets of the spectra in order
to evaluate systematic differences in the effective temperatures,
surface gravities, and iron abundances. We describe them below. Two of
the methods also give the projected rotational velocities, $v \sin i$.

The technique that was applied to all of our TRES, HIRES, and FIES
spectra is referred to as Stellar Parameter Classification (SPC), and
is described fully by \cite{Buchhave:12}. It is based on a
cross-correlation of the observed spectrum against a library of
synthetic spectra calculated from Kurucz model atmospheres
\citep{Castelli:03, Castelli:04}, and determines the temperature,
surface gravity, metallicity, and projected rotational velocity by
seeking the maximum of the cross-correlation coefficient as a function
of those parameters \citep[see also][]{Torres:02, Latham:02}. The grid
of synthetic spectra covers a wide range in the above four parameters,
although it spans only a limited wavelength region between 5050\,\AA\
and 5360\,\AA\ \citep{Buchhave:12}. This allowed the use of five
echelle orders in the FIES spectra, three in the TRES spectra, and two
for HIRES.  For simplicity, all synthetic spectra in this library were
computed with a microturbulent velocity of $\xi = 2$\,\kms\ and a
macroturbulent velocity of $\zeta_{\rm RT} = 1$\,\kms. The
metallicities derived with this method are not strictly [Fe/H], but
represent instead an average abundance [M/H] of the elements producing
absorption features in the spectral region under
consideration. Although iron lines tend to dominate, the two indices
could be different for a star with peculiar abundances, or more
generally for metal-poor stars in which the $\alpha$ elements are
often enhanced. However, for the present sample there is no evidence
of such anomalies from detailed abundance studies by ourselves or
others \citep[e.g.,][]{McCullough:06, Sozzetti:06, Deleuil:08,
Torres:08, Anderson:10a}, nor are any of the stars particularly
metal-poor so that we would expect [$\alpha$/Fe] to be significantly
different from zero. In the following we have therefore assumed that
[M/H] is equivalent to [Fe/H].

A second technique was applied to nearly all of the HIRES spectra, and
relies on the widely used analysis package Spectroscopy Made Easy
\citep[SME; see][]{Valenti:96} with the atomic line database of
\cite{Valenti:05}.  The procedure assumes local thermodynamical
equilibrium (LTE) and plane parallel geometry, and synthesizes spectra
adjusting $T_{\rm eff}$, $\log g$, and [M/H] to achieve the best match
to the observed spectra by minimizing a $\chi^2$ function. [M/H] is a
global abundance parameter in SME that is used to interpolate in the
grid of model atmospheres and to scale the solar abundance pattern
(except for helium and a few other elements including iron) when
calculating opacities. For the present work we report instead the iron
abundance, [Fe/H], which is a separate variable in the fits.  We used
eight wavelength segments of approximately 20\,\AA\ each, one
including the gravity-sensitive Mg\,b triplet and the others spanning
the range 6000--6180\,\AA.  Following \cite{Valenti:05} we used a
fixed value of $\xi = 0.85$\,\kms\ for the microturbulent velocity in
SME, and the radial-tangential macroturbulent velocity was computed
with the prescription given by the same authors.\footnote{A sign in
their formula for macroturbulence was misprinted; the correct
expression is $\zeta_{\rm RT} = 3.98 + (T_{\rm eff}-5770)/650$~\kms.}
The performance of this method for deriving effective temperatures,
surface gravities, metallicities, and rotational velocities is
documented in detail in the same work.

Finally, for about half of the HIRES spectra and some of the TRES
spectra we made use of a third method that follows a more classical
curve-of-growth approach, as implemented in the spectral synthesis
code MOOG\footnote{{\tt http://www.as.utexas.edu/\textasciitilde
chris/moog.html}} \citep{Sneden:73}. We used MOOG in conjunction with
a grid of Kurucz ATLAS model atmospheres \citep{Kurucz:93}. This
technique determines spectroscopic properties under the assumption of
LTE, imposing the conditions of excitation and ionization equilibrium.
It relies on equivalent widths of selected \ion{Fe}{1} and \ion{Fe}{2}
lines, which we measured either manually or using the automated
ARES\footnote{{\tt http://www.astro.up.pt/\textasciitilde
sousasag/ares/}} tool \citep{Sousa:07}. Between 100 and 200 relatively
isolated lines were measured in each spectrum, depending on the SNR,
with equivalent widths in the range from 2 to 120\,m\AA.
Microturbulence was determined by imposing the constraint that the
\ion{Fe}{1} abundance should not depend on the reduced equivalent
width. For details of the procedure see, e.g., \cite{Sozzetti:06,
Sozzetti:07}. For the TRES spectra the wavelength region used is
$\sim$4300--6750\,\AA\ (echelle orders 10 to 38), and for HIRES we
used all orders between 4980\,\AA\ and 6420\,\AA.

\section{Results}
\label{sec:results}

\subsection{Unconstrained determinations}
\label{sec:unconstrained}

In this section we report spectroscopic determinations of the
temperatures, surface gravities, metallicities and projected
rotational velocities with SPC, SME, or MOOG in which we solved for
those three parameters simultaneously, without making use of any
external information about the stars. We refer to these as
``unconstrained'' results, to distinguish them from the results
discussed in the next section in which we hold the surface gravities
fixed.  The unconstrained results are the most commonly reported in
the exoplanet literature, although as we discuss later, the
constrained values are generally preferable. For this reason we defer
a tabulation of the final parameters until
Sect.~\ref{sec:finalresults}.  Tables~\ref{tab:tresstars},
\ref{tab:hiresstars}, and \ref{tab:fiesstars} provide a listing of the
method(s) applied to each spectrum.  For stars with two or more
spectra available from a given telescope we have taken an average of
the individual determinations.

\begin{figure*} 
\epsscale{1.0} 
\plotone{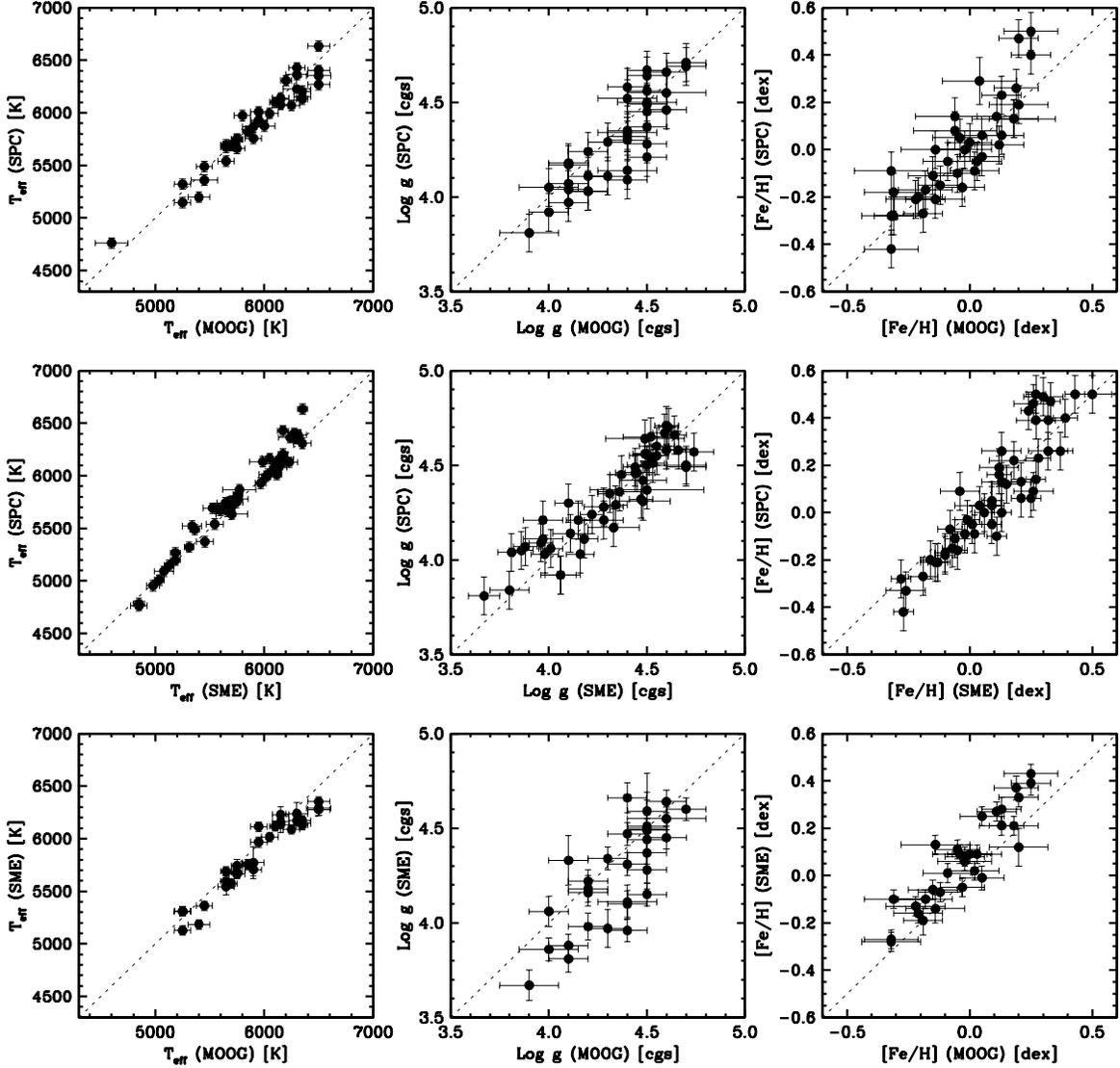}
\figcaption[]{Unconstrained spectroscopic results from SPC, SME, and
MOOG compared against each other for spectra in common (ranging in
number from 31 to 49, from one or more telescopes; see
Table~\ref{tab:dif_unconst}). Parity is indicated by the dotted lines.
\label{fig:unconstrained}} 
\end{figure*} 

A significant fraction of our stars have unconstrained determinations
from two or more analysis methods. This allows us to compare results
and check for systematic differences, which has seldom been done for
transiting planet hosts.  In Figure~\ref{fig:unconstrained} we display
the unconstrained values of $T_{\rm eff}$, $\log g$, and [Fe/H] from
the different procedures against each other. For the purpose of this
comparison the new SME results obtained here have been augmented with
other SME-based results from the literature, relying mostly on HIRES
spectra obtained by the HATNet project \citep{Bakos:04}. Those spectra
have also been reanalyzed with SPC. The average offsets from at least
30 stars in common between the methods are listed in
Table~\ref{tab:dif_unconst}, along with the uncertainty of the
mean. The average differences are well below 100\,K in temperature,
and under 0.1~dex in both $\log g$ and [Fe/H]. In computing these
differences we have adjusted the MOOG abundances to account for a
small difference in the adopted solar iron abundance in our
implementation of that technique ($A({\rm Fe}) =
7.52$)\footnote{$A({\rm Fe}) = \log [N({\rm Fe})/N({\rm H})] + 12$.}
compared to SPC and SME ($A({\rm Fe}) = 7.50$).  On average MOOG is
seen to give slightly hotter temperatures and higher surface gravities
than both SPC and SME, but lower iron abundances.  The SPC results are
intermediate between the other two.

\begin{deluxetable*}{lcccc}
\tablewidth{0pc}
\tablecaption{Comparison of unconstrained results for $T_{\rm eff}$,
$\log g$, and [Fe/H] from different analysis
techniques.\label{tab:dif_unconst}}
\tablehead{
\colhead{} &
\colhead{$\Delta T_{\rm eff}$} &
\colhead{$\Delta \log g$} &
\colhead{$\Delta {\rm [Fe/H]}$} &
\colhead{} \\
\colhead{Methods} &
\colhead{(K)} &
\colhead{(dex)} &
\colhead{(dex)} &
\colhead{$N$}
}
\startdata
SPC$-$SME  & $+32 \pm 12$ & $+0.014 \pm 0.017$ & $-0.023 \pm 0.015$\phantom{\tablenotemark{a}} & 49 \\
SPC$-$MOOG & $-37 \pm 18$ & $-0.049 \pm 0.020$ & $+0.015 \pm 0.018$\tablenotemark{a} & 37 \\
SME$-$MOOG & $-78 \pm 18$ & $-0.093 \pm 0.030$ & $+0.068 \pm 0.014$\tablenotemark{a} & 31 \\ [-2ex]
\enddata
\tablenotetext{a}{MOOG metallicities have been adjusted to the same
solar iron abundance of $A({\rm Fe}) = 7.50$ adopted
in SPC and SME.}
\end{deluxetable*}

\begin{figure} 
\epsscale{1.0} 
\plotone{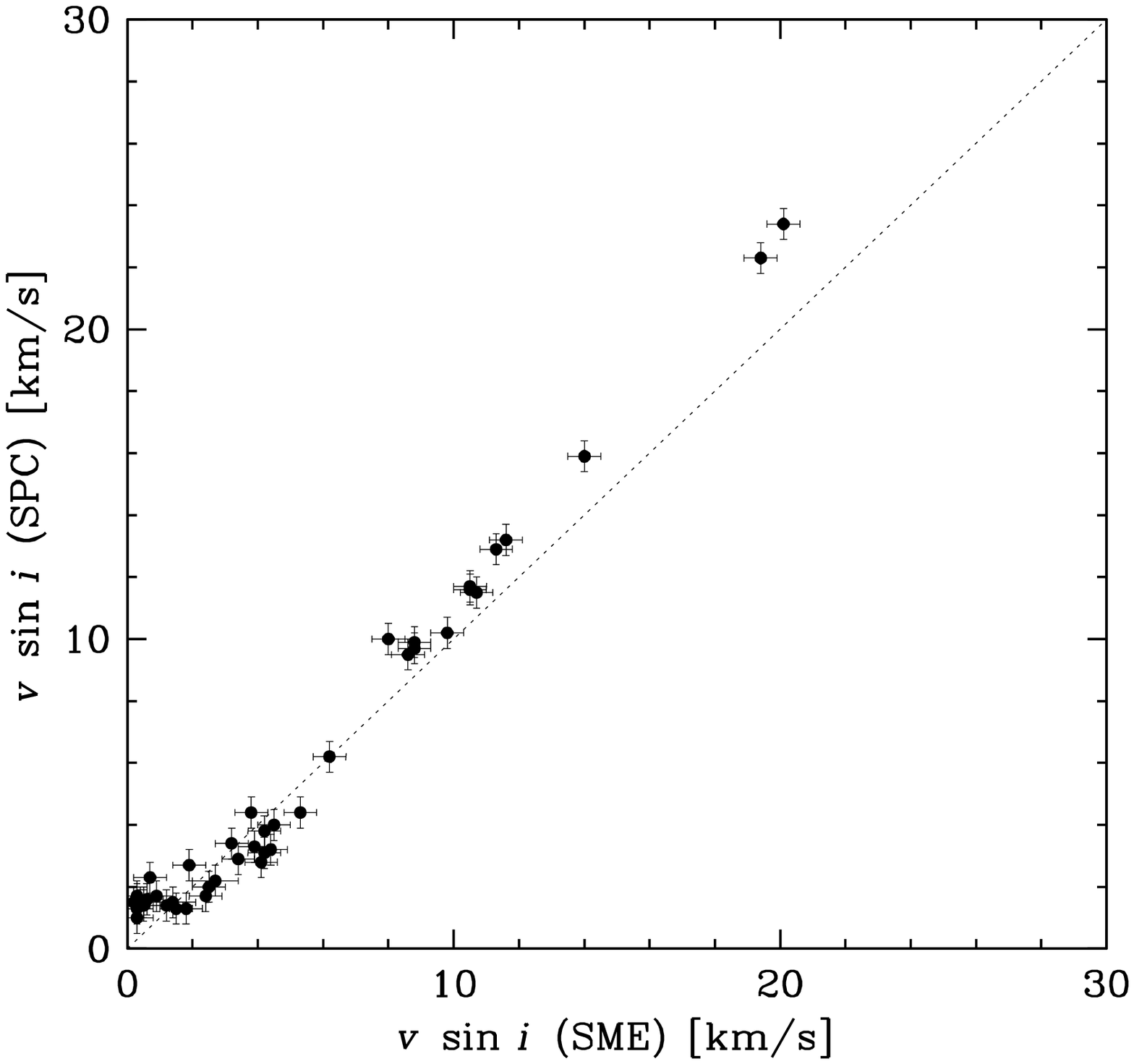}
\figcaption[]{Comparison of $v \sin i$ measurements from SPC and SME,
based on HIRES spectra.\label{fig:vsini}} 
\end{figure} 

A comparison of the projected rotational velocities from SPC and SME
is shown in Figure~\ref{fig:vsini}. For $v \sin i$ values larger than
about 10\,\kms\ the SPC values tend to be systematically larger than
those from SME. Part of this may be related to differences in the
continuum-fitting algorithms used in the two techniques \citep[see,
e.g.,][]{Behr:03, Royer:02a, Royer:02b, Glazunova:08, Hensberge:00}.

\subsection{Constrained determinations}
\label{sec:constrained}

One of the most important uses of the spectroscopic parameters for
transiting planet hosts is to infer the mass and radius of the star
($M_{\star}$, $R_{\star}$), which are needed in turn to establish the
mass and radius of the planet. The stellar dimensions are typically
obtained by comparing $T_{\rm eff}$, $\log g$, and [Fe/H] with stellar
evolution models, or by using empirical calibrations
\citep{Torres:10a, Enoch:10}.  The spectroscopic parameter that most
directly affects the determination of the stellar radius is $\log g$,
which is a proxy for the luminosity (typically unknown for these stars
since the parallaxes have generally not been measured). However,
surface gravity has a rather subtle effect on the spectral line
profiles, and is difficult to measure accurately. It has been
advocated \citep[see][and others]{Sozzetti:07, Holman:07, Torres:08}
that a much better constraint on the luminosity of transiting planet
hosts can be obtained from the normalized semimajor axis $a/R_{\star}$
that is directly measurable from the transit light curve when the
eccentricity is known \citep{Seager:03}. The quantity $a/R_{\star}$ is
closely related to the mean stellar density, $\rho_{\star}$. The
approach that is now most common in the field for inferring
$M_{\star}$ and $R_{\star}$ is to compare $T_{\rm eff}$, [Fe/H], and
$\rho_{\star}$ with stellar evolution models, or to use those three
quantities as inputs to empirical calibrations. Once the mass and
radius are known, a more accurate value of $\log g$ follows trivially.
We refer to this as a density-based surface gravity, which is of
course model- or calibration-dependent.

The $\log g$ values that emerge from this process do not always agree
with the spectroscopic estimates. This inconsistency has usually been
attributed to shortcomings in the spectral synthesis
techniques. Because the surface gravity is typically correlated with
metallicity and temperature in some of the most commonly used methods
of analysis, these differences have motivated some authors to repeat
the spectroscopic determination of $T_{\rm eff}$ and [Fe/H] holding
$\log g$ fixed at the external density-based values, in order to avoid
systematic errors that could propagate into the stellar mass and
radius \citep[see, e.g.,][]{Kovacs:07, McCullough:08}. We have taken
the same approach here, using the best available estimates of the
density-based $\log g$ for each system from published photometric
analyses.  New values of $T_{\rm eff}$, [Fe/H] and $v \sin i$ have
been derived with each of the three methodologies, and are presented
in Table~\ref{tab:constrained}.\footnote{For Kepler-9 and Kepler-11
there are no published determinations of $a/R_{\star}$ or
$\rho_{\star}$ from which to infer a value of $\log g$ that we can use
as a constraint on the spectroscopic determinations. We therefore
exclude these stars from Table~\ref{tab:constrained} and subsequent
discussion.}  The external values of $\log g$ are given later in
Sect.~\ref{sec:finalresults}, with our final results.

\begin{deluxetable*}{lccccccccccc}
\tabletypesize{\scriptsize}
\tablewidth{0pc}
\tablecaption{Spectroscopic results using the external constraint on $\log g$ from photometry (constrained analysis).\label{tab:constrained}}
\tablehead{
& 
\multicolumn{3}{c}{SPC Analysis} &&
\multicolumn{3}{c}{SME Analysis} &&
\multicolumn{2}{c}{MOOG Analysis}  \\ [1.0ex]
\cline{2-4}  \cline{6-8}  \cline{10-11} & \\ [-1.0ex]
\colhead{} &
\colhead{$T_{\rm eff}$} & \colhead{[Fe/H]} & \colhead{$v \sin i$} &&
\colhead{$T_{\rm eff}$} & \colhead{[Fe/H]} & \colhead{$v \sin i$} &&
\colhead{$T_{\rm eff}$} & \colhead{[Fe/H]} & \colhead{} \\
\colhead{Star} & 
\colhead{(K)} & \colhead{(dex)} & \colhead{(\kms)} &&
\colhead{(K)} & \colhead{(dex)} & \colhead{(\kms)} &&
\colhead{(K)} & \colhead{(dex)} & \colhead{Tel\tablenotemark{a}}
}
\startdata
      CoRoT-1 &   $6280 \pm  50$  &  $-0.04 \pm 0.08$  & $  4.3 \pm 0.5$   &&  $6312 \pm  44$ &  $+0.08 \pm 0.04$ &  $  4.9 \pm 0.5$   &&   $6300 \pm  75$  &  $+0.05 \pm 0.11$ &  H \\ 
     CoRoT-2 &   $5552 \pm  50$  &  $-0.15 \pm 0.08$  & $ 10.6 \pm 0.5$   &&  $5602 \pm  44$ &  $-0.01 \pm 0.04$ &  $ 10.0 \pm 0.5$   &&   $5550 \pm  75$  &  $-0.10 \pm 0.11$ &  H \\
     CoRoT-7 &   $5392 \pm  50$  &  $+0.08 \pm 0.08$  & $  0.5 \pm 0.5$   &&  $5274 \pm  44$ &  $+0.02 \pm 0.04$ &  $  0.5 \pm 0.5$   &&   $5250 \pm  75$  &  $ 0.00 \pm 0.08$ &  H \\
   HAT-P-3   &   $5201 \pm  50$  &  $+0.45 \pm 0.08$  & $  1.8 \pm 0.5$   &&     \nodata     &      \nodata      &  $  0.5 \pm 0.5$   &&      \nodata      &      \nodata      &  H \\
   HAT-P-3   &   $5247 \pm  50$  &  $+0.36 \pm 0.08$  & $  2.2 \pm 0.5$   &&     \nodata     &      \nodata      &      \nodata       &&      \nodata      &      \nodata      &  T \\ [-2ex]
\enddata

\tablenotetext{a}{Telescope/instrument combination: H = Keck/HIRES, F
= NOT/FIES, T = FLWO/TRES.}

\tablecomments{Table~\ref{tab:constrained} is published in its
entirety in the electronic edition of \apj. A portion is shown here for
guidance regarding its form and content. Errors are internal.}

\end{deluxetable*}

\begin{figure*}
\epsscale{1.0} 
\plotone{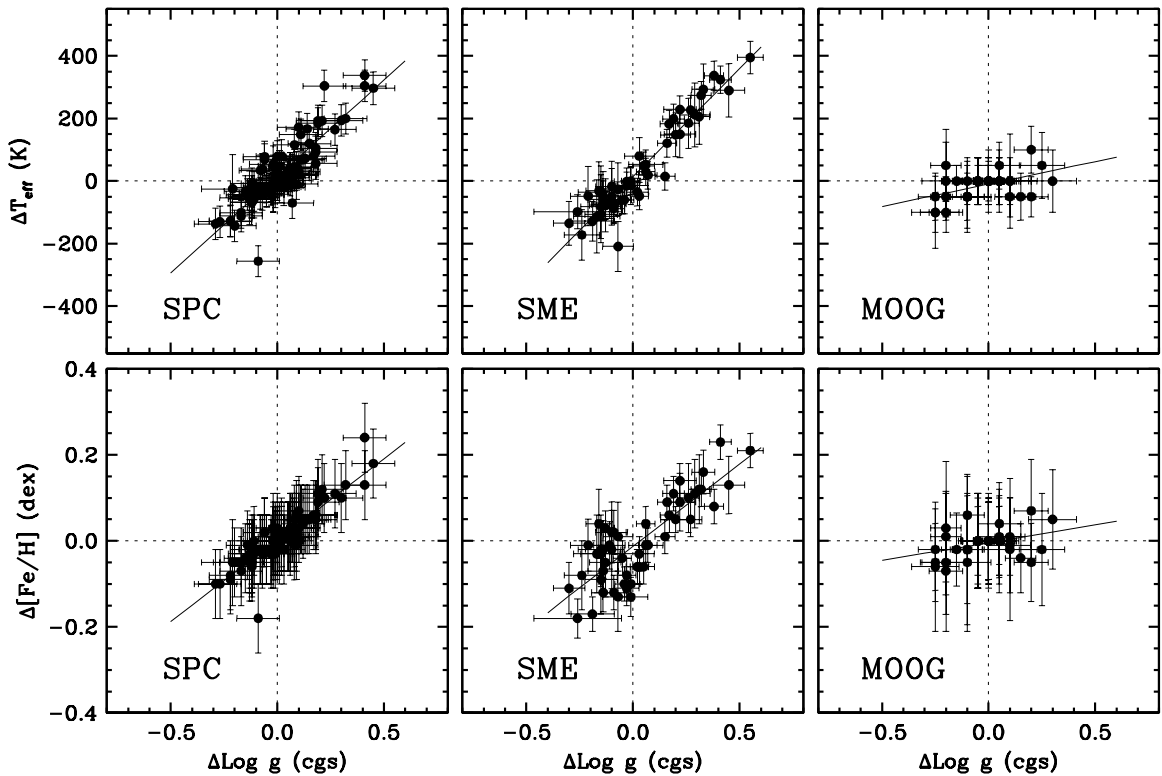}
\figcaption[]{Impact on the temperatures and metallicities of fixing
$\log g$ to the photometric values, for each of the three methods. The
panels show the differences in the sense $\langle$constrained minus
unconstrained$\rangle$ as a function of the change in $\log g$. The
lines represent linear fits computed with the OLS($Y\!\mid\!X$)
procedure as implemented in the {\tt SLOPES} code of
\cite{Feigelson:92}. The Pearson correlation coefficients in $\Delta
T_{\rm eff}$ vs.\ $\Delta\log g$ and $\Delta{\rm [Fe/H]}$ vs.\
$\Delta\log g$ range from 0.82 to 0.95 for SPC and SME, but are only
0.49 and 0.37 for MOOG.
\label{fig:correlations}} 
\end{figure*} 

In Figure~\ref{fig:correlations} we compare the constrained results
against the unconstrained values from the previous section, separately
for the three methods. For SPC and SME the temperature and the
metallicity changes were found to be strongly correlated with changes
in surface gravity, in the sense that $T_{\rm eff}$ and [Fe/H] both
increase when a higher gravity is adopted. The slopes of these
correlations \citep[determined with simple linear regressions of $Y$
on $X$ using the {\tt SLOPES} code of][]{Feigelson:92} are such that
for an increase of 0.5~dex in $\log g$ the temperatures change
systematically by about +310\,K for SPC and +350\,K for SME. For a
similar increase in $\log g$ the metallicities from SPC and SME both
change by about +0.19~dex. These correlations are hardly significant
for MOOG: the formal changes per 0.5~dex increase in surface gravity
are only +70\,K and +0.04~dex in $T_{\rm eff}$ and [Fe/H],
respectively, which are of the order of the typical internal errors or
smaller. We note also that there is some evidence that these
correlations depend on temperature, in the sense that for stars cooler
than the Sun they are roughly half as large.

SPC and SME are somewhat similar techniques in the sense that they
both optimize $T_{\rm eff}$, $\log g$, and [Fe/H] by seeking the best
match between the observed spectrum and a synthetic spectrum. The
correlations described above are not completely unexpected in these
procedures, as one spectroscopic quantity can play against another to
some extent and lead to nearly the same cross-correlation value or
$\chi^2$ value.  For example, stronger lines produced by adopting a
higher metallicity for the synthetic spectrum can be compensated for,
to first order, by a suitable increase in the effective temperature. A
similar degeneracy exists between surface gravity and temperature. In
the case of MOOG the effect of these correlations is evidently much
weaker.
A significant difference between MOOG and the other methods is the
line lists. In particular, for this work MOOG uses only \ion{Fe}{1}
and \ion{Fe}{2} lines, whereas SME additionally includes the region of
the \ion{Mg}{1}\,b triplet, and SPC relies heavily on this same
spectral region as well. Therefore for the two latter methods the
\ion{Mg}{1}\,b lines contain by far the strongest information on $\log
g$. We speculate that any errors in the synthesis of the broad wings
of these pressure-sensitive lines may cause errors in the SPC and SME
determinations that would likely also impact the $T_{\rm eff}$ and
[Fe/H] values because of the correlations described above; MOOG, on
the other hand, would be unaffected. While we have no evidence of such
errors at the present time, this could be investigated by using
different atmospheric models.
We note also that the surface gravity in the unconstrained MOOG
analysis is determined implicitly by requiring that the iron abundance
for \ion{Fe}{1} and \ion{Fe}{2} from the measured equivalent widths be
the same (ionization equilibrium). When fixing $\log g$ to a value
determined externally this condition is generally no longer met,
although the discrepancy for our sample (about 0.05\,dex, on average)
is small compared to the nominal uncertainty in the \ion{Fe}{1}
abundance. This lack of ionization equilibrium seems to have little
effect on the other parameters, and could be a sign of real
differences between \ion{Fe}{1} and \ion{Fe}{2} \citep[see, e.g.,][and
references therein]{Schuler:10}, or deficiencies in the models
\citep{Yong:04}. For all three methods we find that the goodness of
fit (as quantified by $\chi^2$ or the average cross-correlation
coefficient) is only slightly lower in the constrained fits compared
to those in which $\log g$ is left free.

Differences between the results with unconstrained surface gravities
and those using the density-based values were found to be up to
0.5~dex for SPC and SME, and up to 0.3~dex for MOOG. To the extent
that the photometric constraints on $\rho_{\star}$ (and therefore
$\log g$) are accurate, this suggests that all three spectroscopic
procedures are vulnerable to systematics, though perhaps not to the
same degree or for the same reasons.

Indirect evidence of these effects may perhaps be seen already in
published results from the two surveys that have produced the majority
of the transiting planet discoveries from the ground: the WASP project
\citep{Pollacco:06} and the HATNet project \citep{Bakos:04}.  The
HATNet group has generally used SME for the spectroscopic analysis of
the parent stars, and in most cases the studies have been iterated as
described above, using the constraint from the transit light curves to
set $\log g$.  The WASP team has occasionally also used SME, although
more recently they have relied on the UCLSYN package
\citep{Smalley:01} and in general their spectroscopic results are from
an unconstrained analysis ($\log g$ free).

\begin{figure} 
\epsscale{1.15} 
\plotone{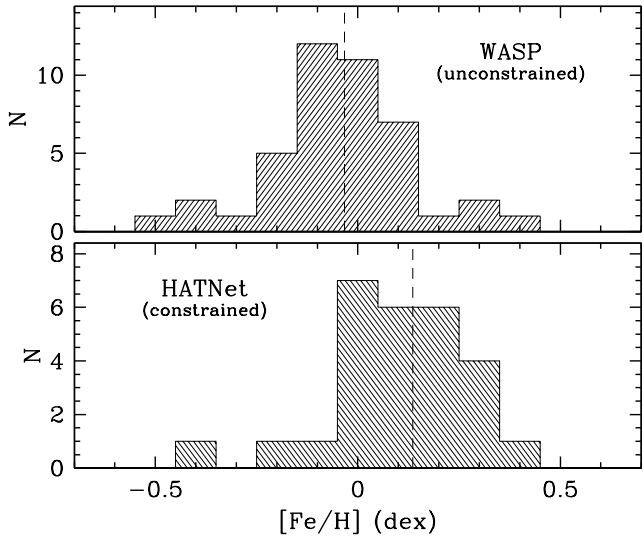}
\figcaption[]{Metallicity distributions for transiting planet hosts
analyzed by the WASP and HATNet groups (43 and 27 stars,
respectively).  The results for the WASP stars are based on
unconstrained spectroscopic analyses, while those from HATNet rely on
$\log g$ as constrained photometrically by the mean stellar
density. The mean of each distribution is indicated with a dashed
line.\label{fig:wasphat}}
\end{figure} 

A comparison of the distribution of published metallicities for the
host stars from these two groups is seen in Figure~\ref{fig:wasphat}.
The average metallicities differ by about 0.17~dex, and a
Kolmogorov-Smirnov test suggests the distributions are statistically
different, with a false alarm probability (FAP) of 0.14\%.  Both
surveys are magnitude limited, with the WASP program being typically
shallower, so a difference in the metallicity distributions is
possible in principle. However, we find that the mean visual magnitude
of the two star samples is essentially the same ($V \approx 11.5$ for
WASP and $V \approx 11.3$ for HATNet), and a K-S test indicates the
brightness distributions are indistinguishable (${\rm FAP} =
0.259$). It seems likely, therefore, that the difference in the
average metallicities is mostly due to the analysis. It goes in the
direction expected, as the constrained $\log g$ values adopted by the
HATNet team are more often larger than the unconstrained values, which
according to Figure~\ref{fig:correlations} should lead to higher
metallicities in that survey, just as observed.

\begin{figure} 
\epsscale{1.15} 
\plotone{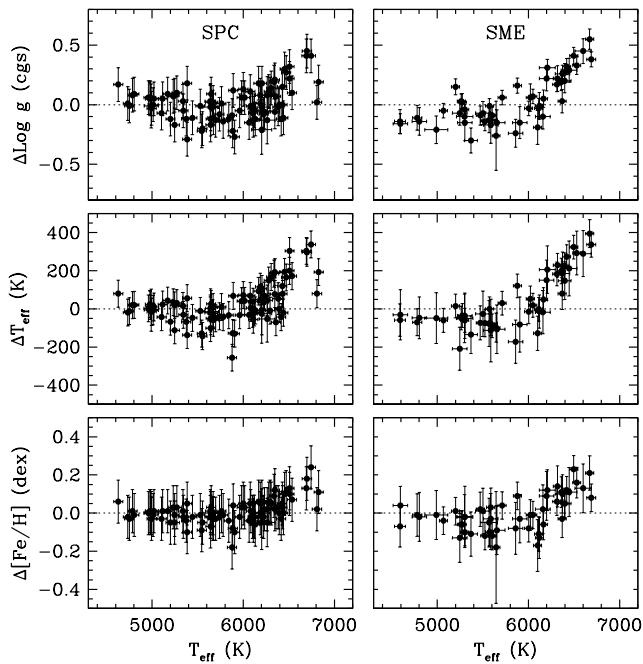}
\figcaption[]{Constrained minus unconstrained differences in surface
gravity, effective temperature, and iron abundance ($\Delta\log g$,
$\Delta T_{\rm eff}$, and $\Delta$[Fe/H]) as a function of effective
temperature, for SPC (left panels) and SME (right).
\label{fig:teff}} 
\end{figure} 

Figure~\ref{fig:teff} shows that for SPC and SME the differences
between the constrained and unconstrained results for $T_{\rm eff}$,
$\log g$, and [Fe/H] are a function of the temperature of the star. In
the case of SME the external surface gravities for stars up to roughly
6000\,K tend to be lower than the unconstrained values, on average,
and in turn lead to systematically lower temperatures and
metallicities. For hotter stars the trend reverses sharply, reaching
maximum differences of 0.5~dex in $\log g$, 400\,K in $T_{\rm eff}$,
and 0.2~dex in metallicity compared to those with $\log g$ free. For
SPC the external $\log g$ values differ somewhat less from the
unconstrained determinations, but again the hotter stars give higher
temperatures and [Fe/H] values when imposing this constraint.  No such
dependence is apparent for the MOOG results. We also found no
significant dependence of $\Delta T_{\rm eff}$, $\Delta \log g$, or
$\Delta {\rm [Fe/H]}$ with either metallicity or surface gravity, in
any of the three methods. The projected rotational velocities from SPC
and SME are insensitive to the change in $\log g$.

As mentioned above the trends in SPC and SME with effective
temperature are most likely related to the fact that the information
on surface gravity in these methods comes almost exclusively from the
pressure-broadened wings of the \ion{Mg}{1}\,b lines at
$\sim$5200\,\AA.  For stars hotter than about 6000\,K the wings of
these lines weaken considerably, and the methods lose sensitivity to
$\log g$.  It is less clear why the bias in $\log g$ for such stars is
toward smaller values, as opposed to being toward higher values or
simply having a larger scatter.  In any case, the positive correlation
between $\log g$, $T_{\rm eff}$, and [Fe/H] explains the upturn in the
last two quantities in Figure~\ref{fig:teff}.

\begin{figure} 
\epsscale{1.15}
\plotone{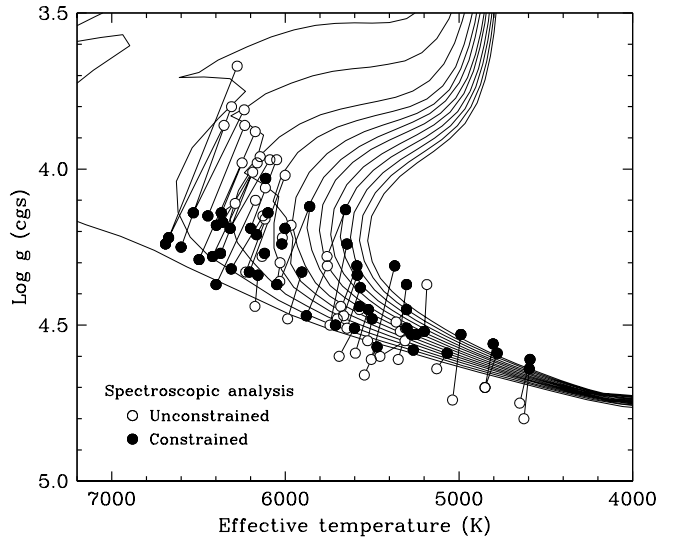}
\figcaption[]{Illustration of the impact that constraining $\log g$
has on the location of transiting planet host stars in the H-R
diagram, shown here for spectra analyzed with SME. We include
additional stars with SME-based results from the literature carried
out in the same way as in this work.  Unconstrained and constrained
results are connected with lines, and shown against representative
stellar evolution models from \cite{Yi:01} for solar metallicity, and
ages between 1 and 13~Gyr (in steps of 1~Gyr).\label{fig:hrsme}}
\end{figure}

As a result of the temperature correlation for SPC and SME shown in
Figure~\ref{fig:teff}, the net effect of applying the $\log g$
constraint with those methods is to shift the stars in the H-R diagram
towards a less evolved state (closer to the zero-age main sequence).
The less evolved state also seems more likely \emph{a priori}, because
of the slower speed of evolution for unevolved stars. We illustrate
this for SME in Figure~\ref{fig:hrsme}.  The trend has a significant
impact on the inferred radii of the stars, which we discuss more
quantitatively in Sect.~\ref{sec:impact}.

\begin{figure}
\epsscale{1.15} 
\plotone{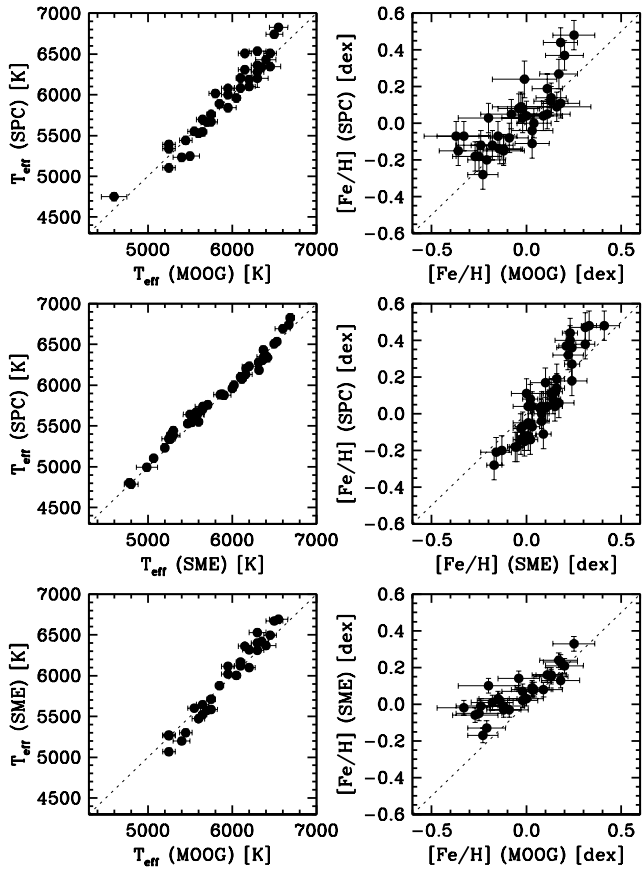}
\figcaption[]{Constrained spectroscopic results from SPC, SME, and
MOOG compared against each other for spectra in common (ranging from
29 to 44, from one or more telescopes; see Table~\ref{tab:dif_const}).
One-to-one relations are indicated with dotted lines.
\label{fig:constrained}} 
\end{figure}

The constrained results from the different methods are compared in
Figure~\ref{fig:constrained} for stars in common.  Our expectation was
that the methods would show better agreement than the unconstrained
results, since some of the biases in the determination of $T_{\rm
eff}$ and [Fe/H] that are intrinsic to each of the techniques would be
reduced by enforcing an external constraint on $\log g$.  This is
indeed the case, although residual trends can still be seen in some of
the panels of Figure~\ref{fig:constrained}, such as between MOOG and
SME in [Fe/H], or between SPC and SME in temperature and also
metallicity. These are most likely a reflection of strong correlations
between the spectroscopic parameters that are present in at least two
of the methods (SPC and SME), as already described, and the different
degrees to which the three procedures respond to the imposition of an
external $\log g$ constraint. Many details of the spectroscopic
analysis are likely to influence the results in ways that are
difficult to predict or quantify. This includes the accuracy of the
continuum normalization of the spectra, the model atmospheres used,
the adopted line lists, and the sensitivity to the SNR of the spectra,
which varies from method to method.  Nevertheless, average differences
between the final values from SPC, SME, and MOOG remain quite small,
as seen in Table~\ref{tab:dif_const}.

\begin{deluxetable}{lccc}
\tablewidth{0pc}
\tablecaption{Comparison of constrained results for $T_{\rm eff}$
and [Fe/H] from different analysis techniques.\label{tab:dif_const}}
\tablehead{
\colhead{} &
\colhead{$\Delta T_{\rm eff}$} &
\colhead{$\Delta {\rm [Fe/H]}$} &
\colhead{} \\
\colhead{Methods} &
\colhead{(K)} &
\colhead{(dex)} &
\colhead{$N$}
}
\startdata
SPC$-$SME  & $+27 \pm 9$\phn &  $-0.020 \pm 0.015$\phantom{\tablenotemark{a}} & 44 \\
SPC$-$MOOG & $+27 \pm 23$    &  $+0.049 \pm 0.019$\tablenotemark{a} & 36 \\
SME$-$MOOG & \phn$+8 \pm 22$ &  $+0.081 \pm 0.017$\tablenotemark{a} & 29 \\ [-2ex]
\enddata
\tablenotetext{a}{MOOG metallicities have been adjusted to the same
solar iron abundance of $A({\rm Fe}) = 7.50$ adopted in SPC and SME.}
\end{deluxetable}

\subsection{Final spectroscopic results}
\label{sec:finalresults}

The sensitivity of SPC and SME to changes in the surface gravity
suggests these techniques are more susceptible to systematic bias than
MOOG, although $\log g$ errors up to 0.3~dex occur even with MOOG. The
smaller impact of those errors on $T_{\rm eff}$ and [Fe/H] in the
latter method does not necessarily guarantee those quantities are free
from bias. It is difficult to ascertain whether the results from one
technique are more accurate than another. Independent checks on the
temperatures could in principle be obtained from the measurement of
bolometric fluxes and precise interferometric angular diameters, but
the stars in the present sample are generally too faint for current
long-baseline interferometers. Photometric temperature estimates based
on color indices might provide an alternative, but suffer from the
possibility of reddening, which is unknown for most of these stars.

Given that the average systematic differences in
Table~\ref{tab:dif_const} between the three methodologies are
relatively small, the final spectroscopic parameters we adopted for
stars analyzed with more than one method or observed using more than
one instrument is the weighted mean, which should not significantly
affect the homogeneity of the results. We collect the final parameters
for 56 transiting planet hosts in Table~\ref{tab:finalresults}, where
the uncertainties include a contribution added in quadrature from the
overall scatter of the measurements available for stars with multiple
determinations.  These added dispersions are $\sigma_{T_{\rm eff}} =
59$\,K, $\sigma_{\rm [Fe/H]} = 0.062$\,dex, and $\sigma_{v \sin i} =
0.85$\,\kms. The table also reports the external constraint on $\log
g$ used for the second iteration of the spectroscopic analyses with
SPC, SME, and MOOG.

\begin{deluxetable*}{lcccccc}

\tablecaption{Final spectroscopic results (average of all new
determinations).\label{tab:finalresults}}

\tablehead{
\colhead{} &
\colhead{$T_{\rm eff}$} &
\colhead{${\rm [Fe/H]}$} &
\colhead{$v \sin i$} &
\colhead{} &
\colhead{External} &
\colhead{}
\\
\colhead{Name} &
\colhead{(K)} &
\colhead{(dex)} &
\colhead{(\kms)} &
\colhead{$N$\tablenotemark{a}} &
\colhead{$\log g$ (cgs)} &
\colhead{Source\tablenotemark{b}}
}
\startdata
      CoRoT-1  &  $6298 \pm  66$  &  $+0.06 \pm 0.07$  & $ 4.6 \pm 0.9$  &  3,2  &  $4.33 \pm 0.01$  &       1     \\         
      CoRoT-2  &  $5575 \pm  66$  &  $-0.04 \pm 0.08$  & $10.3 \pm 0.9$  &  3,2  &  $4.51 \pm 0.04$  &       2     \\ 	 
      CoRoT-7  &  $5313 \pm  73$  &  $+0.03 \pm 0.07$  & $ 0.5 \pm 1.0$  &  3,1  &  $4.54 \pm 0.04$  &       3     \\ 	 
      HAT-P-3  &  $5224 \pm  69$  &  $+0.41 \pm 0.08$  & $ 1.5 \pm 1.0$  &  2,3  &  $4.58 \pm 0.03$  &       4     \\ 	 
      HAT-P-4  &  $5890 \pm  67$  &  $+0.20 \pm 0.08$  & $ 5.6 \pm 0.9$  &  3,3  &  $4.14 \pm 0.03$  &       5    \\ [-2ex] 	 
\enddata

\tablenotetext{a}{Number of determinations for $T_{\rm eff}$ and
[Fe/H] (based on spectra from different telescopes, or derived with
different methodologies), followed by the number of determinations for
$v \sin i$.}

\tablenotetext{b}{Sources for the external $\log g$ constraint (either
reported directly in these studies, or inferred from $M_{\star}$ and
$R_{\star}$ as reported there, or based on the reported $a/R_{\star}$
values and our own stellar evolution modeling using isochrones from
\citealt{Yi:01}). References: 
(1) \cite{Bean:09}; 
(2) \cite{Gillon:10}; 
(3) \cite{Leger:09}; 
(4) \cite{Torres:07}; 
(5) \cite{Kovacs:07}; 
(6) \cite{Bakos:07}; 
(7) \cite{Noyes:08}; 
(8) \cite{Christensen-Dalsgaard:10}; 
(9) \cite{Latham:09}; 
(10) \cite{Shporer:09}; 
(11) \cite{West:09}; 
(12) \cite{Bakos:10}; 
(13) \cite{Bakos:09}; 
(14) \cite{Torres:10b}; 
(15) \cite{Kovacs:10}; 
(16) \cite{Buchhave:10b}; 
(17) \cite{Howard:12}; 
(18) \cite{Hartman:11a};
(19) \cite{Bakos:11}; 
(20) \cite{Kipping:10}; 
(21) \cite{Quinn:12};
(22) \cite{Hartman:11b}; 
(23) \cite{Buchhave:11}; 
(24) \cite{Gilliland:11}; 
(25) \cite{Hebrard:10}; 
(26) \cite{Pal:10}; 
(27) \cite{Torres:08}; 
(28) \cite{Kipping:11a};
(29) \cite{Batalha:11};
(30) \cite{Kipping:11b};
(31) \cite{Sozzetti:09};
(32) \cite{Simpson:10};
(33) \cite{Christian:09};
(34) \cite{Enoch:10};
(35) \cite{Skillen:09};
(36) \cite{Joshi:09};
(37) \cite{Anderson:10a}
(38) \cite{Southworth:09b};
(39) \cite{Anderson:10b};
(40) \cite{Street:10};
(41) \cite{Anderson:11};
(42) \cite{Barros:11};
(43) \cite{Burke:10};
(44) \cite{Fernandez:09};
(45) \cite{Winn:08};
(46) \cite{Narita:10}.
}

\tablecomments{Table~\ref{tab:finalresults} is published in its entirety in
the electronic edition of \apj. A portion is shown here for guidance
regarding its form and content.}

\end{deluxetable*}

\section{Comparison with other determinations}
\label{sec:comparison}

Prior to this study, the largest effort to determine spectroscopic
parameters for transiting planet hosts in a homogeneous way was that
of \cite{Ammler:09}, who used MOOG following the precepts of
\cite{Santos:04, Santos:06}. Our sample and theirs have 14 stars in
common.

\begin{figure}
\epsscale{1.15} 
\plotone{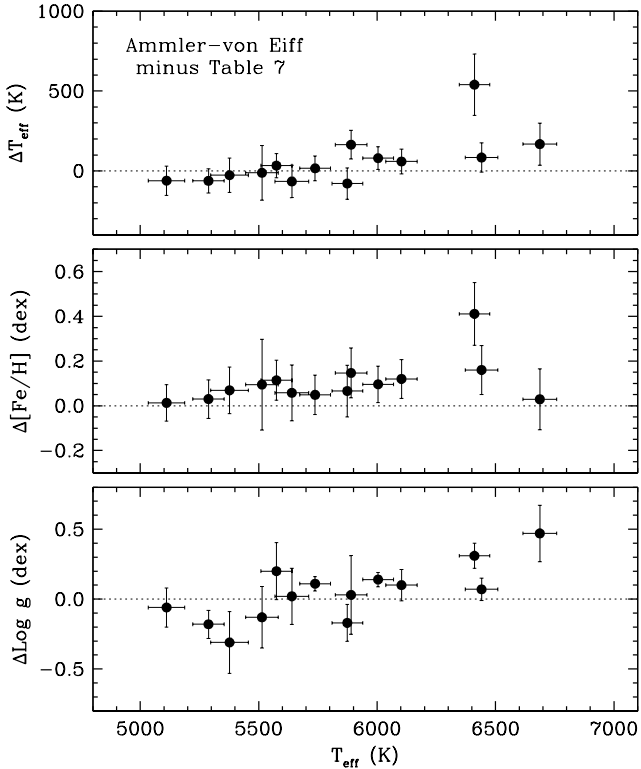}
\figcaption[]{Comparison of our results as listed in
Table~\ref{tab:finalresults} with those of \cite{Ammler:09}, for the
14 stars in common. The temperature, metallicity, and gravity
differences are in the sense ``theirs minus ours'', where our $\log g$
values are the density-based determinations. The temperatures plotted
on the horizontal scale are our own. The high value in the top two
panels corresponds to HD~147506 (HAT-P-2).
\label{fig:ammler1}} 
\end{figure}

A comparison of our results from Table~\ref{tab:finalresults} with
theirs is seen in Figure~\ref{fig:ammler1}, where the differences in
$T_{\rm eff}$, [Fe/H], and $\log g$ for these 14 stars are plotted in
the sense $\langle$Ammler-von Eiff minus
Table~\ref{tab:finalresults}$\rangle$. As indicated earlier our
adopted surface gravities are the density-based values, not the
spectroscopic values. The temperatures on the horizontal axis of the
figure are from our own determinations.

Although the sample is small, there appears to be a significant
systematic difference in the surface gravities that correlates with
temperature (bottom panel), such that \cite{Ammler:09} measure a
higher $\log g$ for the hotter stars.  It is difficult to see how a
bias in our $\log g$ values from Table~\ref{tab:finalresults} could
cause this, as those determinations rest essentially on the
$a/R_{\star}$ measurements from the transit light curves.  There may
also be a trend of $\Delta T_{\rm eff}$ with temperature (top panel),
though it seems marginal. For the metallicities there is no
significant correlation with temperature, but we note that \emph{all}
of the \cite{Ammler:09} values of [Fe/H] are higher than ours, the
average difference being $+0.10 \pm 0.03$~dex.

\section{Impact of the surface gravity constraint on stellar masses and radii}
\label{sec:impact}

In Figure~\ref{fig:hrsme} above we presented a graphical illustration
of how the inferred evolutionary state of the star can change
considerably as a result of applying the external (photometric)
constraint on $\log g$ in the spectroscopic analysis. As might be
expected, this can also lead to significant changes in the mass and
radius derived for the star, which are of more immediate interest for
computing planetary properties in transiting planet systems.

To explore this more quantitatively we have derived stellar masses and
radii for the same stars shown in Figure~\ref{fig:hrsme} following the
common procedure of comparing the spectroscopic quantities $T_{\rm
eff}$, $\log g$, and [Fe/H] against stellar evolution models.  For
each star we performed a Monte Carlo simulation in which we drew
10,000 values of the three spectroscopic parameters from Gaussian
distributions characterized by the measured values and observational
errors, assuming they are uncorrelated. We compared each set with
isochrones from the Yonsei-Yale series by \cite{Yi:01}, seeking the
best match in a $\chi^2$ sense. Ages were allowed to vary in steps of
0.1~Gyr over the range from 0.1 to 13.7~Gyr, and the $\alpha$-element
abundance was assumed to be solar for this test ([$\alpha$/Fe] =
0.0). Stellar masses and radii were determined from the mode of the
respective posterior probability distributions, and 1-$\sigma$ lower
and upper confidence limits were defined by the 15.85\% and 84.15\%
percentiles of the cumulative distribution. We carried out these
calculations first using the unconstrained SME results, and then
repeated the process with the constrained results.

\begin{figure}
\epsscale{1.15}
\plotone{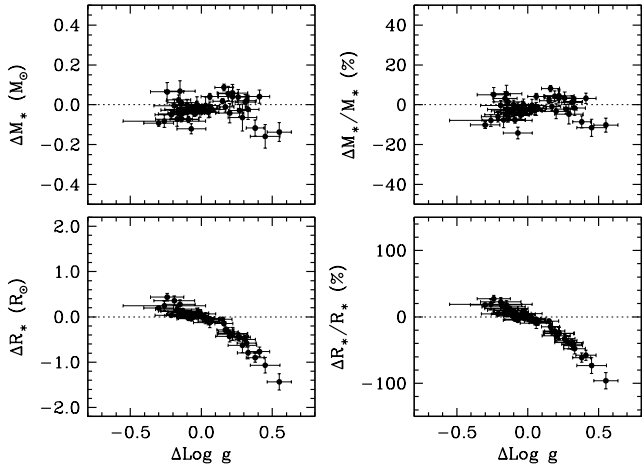}
\figcaption[]{Mass and radius differences resulting from the use of
constrained and unconstrained spectroscopic properties from SME along
with stellar evolution models. Differences in the sense
$\langle$constrained minus unconstrained$\rangle$ are shown in
absolute units on the left, and as a percentage of $M_{\star}$ or
$R_{\star}$ on the right.\label{fig:compareMRuc1}}

\end{figure}

A comparison between $M_{\star}$ and $R_{\star}$ from these two sets
of parameters is shown in Figure~\ref{fig:compareMRuc1} as a function
of $\Delta\log g$, where the changes are displayed both in absolute
units and as a percentage.  In the most extreme cases the
unconstrained radii can be off by up to 100\%.  This occurs for the
hotter stars, as can be seen in Figure~\ref{fig:hrsme}.  The impact on
the mass is also not negligible, reaching $\sim$15\% in some cases.
Errors of this magnitude are almost always larger than other
observational errors.  Very similar results were obtained using the
SPC determinations.

The importance of this bias coming from the reliance on the weakly
determined spectroscopic $\log g$ is now widely recognized in the
community, and is largely avoided in current analyses of transiting
planets by adopting instead the density-based $\log g$ to infer
$M_{\star}$ and $R_{\star}$.  However, many of those same studies
still retain the temperatures and metallicities from
\emph{unconstrained} spectroscopic analyses in which the surface
gravity was left free.  Because $T_{\rm eff}$ and [Fe/H] typically
suffer from strong correlations, as shown earlier in
Figure~\ref{fig:correlations}, those values have a residual bias that
is often not negligible and can propagate to the stellar masses and
radii. This bias has been generally overlooked.

\begin{figure}[t]
\epsscale{1.15}
\plotone{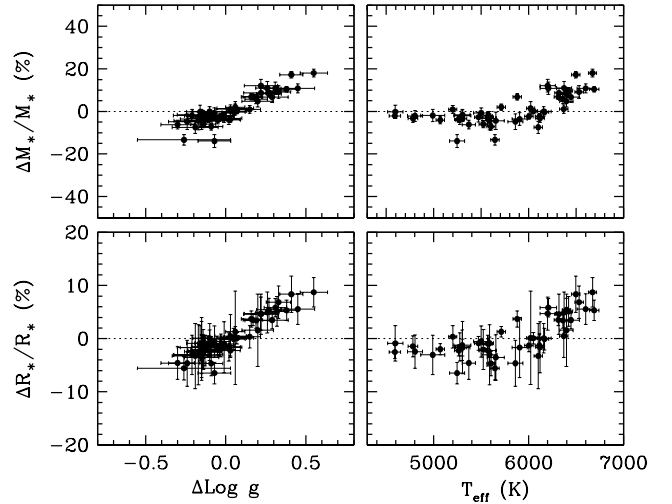}
\figcaption[]{Systematic errors in the stellar mass and radius
(expressed as a percentage) when using \emph{unconstrained} values of
$T_{\rm eff}$ and [Fe/H] from SME together with the external
photometric constraint on $\log g$ from the mean stellar density. The
differences shown are between the mixed usage just mentioned and the
constrained results from a second iteration of SME described in the
text, in the sense $\langle$mixed minus constrained$\rangle$.
\label{fig:compareMRmc1}}
\end{figure}

We have quantified this systematic effect by repeating the
determination of $M_{\star}$ and $R_{\star}$ for the same sample above
using the photometry-based $\log g$, but deliberately adopting $T_{\rm
eff}$ and [Fe/H] from the unconstrained spectroscopic analysis, to
emulate the procedure often followed in published studies.
Figure~\ref{fig:compareMRmc1} compares these masses and radii with
those based on our second iteration of SME, in which the temperature
and metallicity were redetermined using the external $\log g$
constraint.  There are clear differences in the stellar properties
that mimic those seen in Figure~\ref{fig:teff}.  We conclude that even
if a more accurate $\log g$ is adopted for inferring $M_{\star}$ and
$R_{\star}$ from evolutionary models, using the unconstrained values
of the temperatures and metallicities rather than those from a second
iteration of the spectroscopic analysis can still lead to stellar
masses that are up to 20\% too large, and radii that are overestimated
by as much as 10\% for the hotter stars. Since planetary properties
from Doppler and light-curve analyses have dependencies $M_p \propto
M_{\star}^{2/3}$ and $R_p \propto R_{\star}$, the above errors can
translate into biases (overestimates) of 13\% and 10\% in the
planetary masses and radii.  For cooler stars the errors tend to be in
the opposite direction, and are smaller. Analogous results were
obtained when using the determinations from SPC.  In the case of MOOG,
on the other hand, the effect on $M_{\star}$ and $R_{\star}$ from
using unconstrained temperatures and metallicities is only marginally
significant, consistent with the much smaller correlations seen in
Figure~\ref{fig:correlations}.

We defer a comprehensive examination and tabulation of stellar masses
and radii for a future paper, since we are also compiling many new
light curves for analysis, which may improve the external constraints.

\section{Discussion}
\label{sec:discussion}

While most transiting planet investigations use stellar evolution
models to infer stellar masses and radii, others use empirical
relations for $M_{\star}$ and $R_{\star}$ as a function of
temperature, metallicity, and mean density. The biases described above
can affect both.  The {\it Kepler\/} Mission is increasingly making
use of asteroseismology as an alternate way of deriving the mean
stellar density, based on oscillation frequencies measured directly
from the light curves in favorable cases. Deriving the stellar mass or
radius with this technique still requires stellar models, as well as
accurate measurements of $T_{\rm eff}$ and [Fe/H], so there is still a
danger of systematic errors from the use of unconstrained
spectroscopic determinations.

Numerous investigations have addressed the persistent problem of the
anomalously ``inflated'' radii of some of the Jovian planets ---
objects that can exceed 60\% of the size of Jupiter in extreme cases
--- which has been with us since the discovery of transits in
HD~209458, challenging our understanding of planet formation and
evolution.  A variety of mechanisms have been proposed that may play a
role in ``puffing up'' the planets, but no universal process seems to
account for all of these anomalies \citep[see, e.g.,][and references
therein]{Burrows:07, Miller:09, Fortney:10, Demory:11}. Our results in
the preceding section suggest that a portion of the discrepancy in
$R_p$ may have to do with systematic errors in the stellar radii in
some cases, which can amount to $\sim$10\%, as illustrated in
Figure~\ref{fig:compareMRmc1}.

One of the notorious examples of inflated planets for which we have
derived new spectroscopic parameters is WASP-12\,b, with a radius of
$R_p = 1.79 \pm 0.09\,R_{\rm Jup}$
\citep{Hebb:09}.\footnote{\cite{Chan:11} have recently revised this
  value downward slightly to $R_p = 1.736 \pm 0.092\,R_{\rm Jup}$,
  adopting the same spectroscopic parameters as the original authors.}
This study of WASP-12\,b adopted spectroscopic properties derived from
an unconstrained SME analysis yielding $T_{\rm eff} = 6290$\,K
(rounded off by the authors to 6300\,K, and assigned errors of
$^{+200}_{-100}$\,K) and ${\rm [M/H]} = +0.30^{+0.05}_{-0.15}$. These
are both somewhat higher than we derive here from our constrained
spectroscopic analyses: $T_{\rm eff} = 6118 \pm 64$\,K and ${\rm
  [Fe/H]} = +0.07 \pm 0.07$. We note that their spectroscopic surface
gravity ($\log g = 4.38$) is also larger than the density-based
(photometrically constrained) value they reported, $\log g = 4.17 \pm
0.03$. These differences are all consistent with the magnitude and
sign of correlations shown in Figure~\ref{fig:correlations}, and leave
open the possibility that the stellar mass and radius of
\cite{Hebb:09} may be biased. To test this we repeated their isochrone
analysis using the Yonsei-Yale stellar evolution models of
\cite{Yi:01}, first with the \cite{Hebb:09} temperature and
metallicity, and then adopting ours. The surface gravity was held
fixed at the density-based value they determined. The stellar radius
we obtained with our revised spectroscopic parameters is 5\% smaller,
implying a planetary radius also 5\% smaller, all else being
equal. While this correction is far from what would be needed to solve
the puzzle of the inflated radius of WASP-12\,b, it does go in the
right direction.

\section{Conclusions}
\label{sec:conclusions}

Accurate knowledge of the properties of the host stars in transiting
exoplanet systems is essential to derive accurate
characteristics for the planets. Considerable efforts have been
devoted to improving the light curves of newly discovered as well as
previously known transiting systems, but relatively little attention
has been paid to refining the spectroscopic properties of the stars.

Here we have derived new effective temperatures, metallicities, and
projected rotational velocities for 56 transiting planet systems in a
homogeneous manner, using the SPC technique.  These determinations
bring a needed measure of uniformity to the growing collection of
stellar and planetary properties that should facilitate the discovery
of patterns and correlations that may provide valuable insight into
the nature of planets.

A key aspect of our spectral analysis is the application of an
external constraint on the surface gravity, based on accurate
knowledge of the mean stellar density of the star, which comes
directly from the light curve modeling. Because $\log g$ is usually
weakly constrained by the spectra, fixing it as we have done here
prevents errors in $\log g$ from biasing the temperatures and
metallicities, and from affecting the inferred stellar masses and
radii.  Those biases come mainly from strong correlations between
$T_{\rm eff}$, [Fe/H], and $\log g$ that are present in unconstrained
determinations ($\log g$ free), not only when applying SPC, but also
in the widely used SME procedure as implemented by \cite{Valenti:05}.
Both of these methods are based on spectral synthesis. We find that
the correlations are much smaller with MOOG, which uses a more
classical curve-of-growth approach.

We investigate the interagreement among the three spectroscopic
techniques by applying SME and MOOG to subsets of our stars, and we
show that the temperatures and metallicities are generally in good
accord after application of the $\log g$ constraint (the mean
differences being well under 50\,K and 0.1~dex, respectively). We do,
however, detect some remaining systematic trends as a function of
temperature and metallicity that are occasionally larger in some
regimes.

Virtually all current studies of transiting planets make use of the
mean stellar density as a luminosity indicator to derive the stellar
properties, either through a comparison with model isochrones, or
using empirical relations. We demonstrate that not using
$\rho_{\star}$ can incur errors in mass of up to 20\%, and errors in
the radius as large as 100\% in some cases. Even though such errors
are now usually avoided, many authors still retain the temperatures
and metallicities obtained from \emph{unconstrained} spectroscopic
analyses, i.e., without fixing $\log g$ to the more accurate values
based on the light curve modeling. We demonstrate that this practice
can lead to residual biases in $M_{\star}$ of up to 20\%, and
systematic errors in $R_{\star}$ up to 10\% for the hotter stars,
which will propagate through to the planetary properties. Such errors
can be larger than other observational uncertainties, and may explain
part of radius anomaly of some of the inflated Jovian planets.  In
order to avoid this, we advocate performing a second iteration on the
spectroscopic analysis (particularly when using SPC or SME) that fixes
$\log g$ to the value inferred from the photometrically determined
mean stellar density.

\acknowledgements

We are grateful to Jeff Valenti for helpful discussions, and to 
P.\ Berlind,
M.\ Calkins,
G.\ Esquerdo,
D.\ W.\ Latham, and
R.\ P.\ Stefanik
for their help in obtaining the TRES spectra used here.  We also thank
the anonymous referee for helpful comments. GT acknowledges partial
support for this work from NASA's Origins of Solar Systems program,
through grant NNX09AF59G, and from NSF grant AST-10-07992.  The
research has made use of the SIMBAD database, operated at CDS,
Strasbourg, France, and of NASA's Astrophysics Data System Abstract
Service.



\begin{thebibliography}{}

\bibitem[Ammler-von Eiff et al.(2009)]{Ammler:09}
 Ammler-von Eiff, M.\ et al.\ 2009, \aap, 507, 523

\bibitem[Andersen(1991)]{Andersen:91}
 Andersen, J. 1991, \aapr, 3, 91

\bibitem[Anderson et al.(2010a)]{Anderson:10a}
 Anderson, D.\ R.\ 2010a, \apj, 709, 159

 \bibitem[Anderson et al.(2010b)]{Anderson:10b} Anderson,
D.\ R., Gillon, M., Maxted, P.\ F.\ L., et al.\ 2010b, \aap, 513, L3 

\bibitem[Anderson et al.(2011)]{Anderson:11} Anderson,
D.\ R., Collier Cameron, A., Hellier, C., et al.\ 2011, \aap, 531, A60 

\bibitem[Bakos et al.(2004)]{Bakos:04}
 Bakos, G., Noyes, R.\ W., Kov\'acs, G., Stanek, K.\ Z., Sasselov, D.\
 D., \& Domsa, I. 2004, \pasp, 116, 266

\bibitem[Bakos et al.(2007)]{Bakos:07} Bakos,
G.\ {\'A}., Shporer, A., P{\'a}l, A., et al.\ 2007, \apjl, 671, L173 

\bibitem[Bakos et al.(2010)]{Bakos:10} Bakos,
G.\ {\'A}., Torres, G., P{\'a}l, A., et al.\ 2010, \apj, 710, 1724 

\bibitem[Bakos et al.(2009)]{Bakos:09} Bakos,
G.\ {\'A}., Howard, A.\ W., Noyes, R.\ W., et al.\ 2009, \apj, 707, 446 

\bibitem[Bakos et al.(2011)]{Bakos:11} Bakos,
G.\ {\'A}., Hartman, J., Torres, G., et al.\ 2011, \apj, 742, 116 

\bibitem[Barros et al.(2011)]{Barros:11} Barros,
S.\ C.\ C., Faedi, F., Collier Cameron, A., et al.\ 2011, \aap, 525, A54 

\bibitem[Batalha et al.(2011)]{Batalha:11} Batalha,
N.\ M., Borucki, W.\ J., Bryson, S.\ T., et al.\ 2011, \apj, 729, 27 

\bibitem[Bean(2009)]{Bean:09} Bean, J.\ L.\ 2009, \aap, 506, 369 

\bibitem[Behr(2003)]{Behr:03} Behr, B.\ B. 2003, \apjs, 149, 101

\bibitem[Buchhave et al.(2010a)]{Buchhave:10a}
 Buchhave, L.\ A. 2010a, Ph.D.\ thesis, Univ.\ of Copenhagen, \hfill\linebreak
{\tt http://intersigma.dk/data/thesis/Thesis\_Lars\_Buchhave.pdf}

\bibitem[Buchhave et al.(2010b)]{Buchhave:10b}
 Buchhave, L.\ A.\ et al.\ 2010b, \apj, 720, 1118

\bibitem[Buchhave et al.(2012)]{Buchhave:12}
 Buchhave, L.\ A.\ et al.\ 2012, \nat, 486, 375

\bibitem[Buchhave et al.(2011)]{Buchhave:11}
 Buchhave, L.\ A., Bakos, G.\ {\'A}., Hartman, J.\ D., et al.\ 2011,
\apj, 733, 116 

\bibitem[Burke et al.(2010)]{Burke:10} Burke, C.\ J.,
McCullough, P.\ R., Bergeron, L.\ E., et al.\ 2010, \apj, 719, 1796 

\bibitem[Burrows et al.(2007)]{Burrows:07}
 Burrows, A., Hubeny, I., Budaj, J., \& Hubbard, W.\ B. 2007, \apj,
 661, 502

\bibitem[Castelli \& Kurucz(2003)]{Castelli:03}
 Castelli, F., \& Kurucz, R.\ L. 2003, in Modelling of Stellar
 Atmospheres, IAU Symp.\ 210, eds.\ N.\ Piskunov, W.\ W.\ Weiss, and
 D.\ F.\ Gray (San Francisco: ASP), p.\ A20

\bibitem[Castelli \& Kurucz(2004)]{Castelli:04}
 Castelli, F., \& Kurucz, R.\ L. 2004, arXiv:astro-ph/0405087

\bibitem[Chan et al.(2011)]{Chan:11}
 Chan, T., Ingemyr, M., Winn, J.\ N., Holman, M.\ J., Sanchis-Ojeda,
 R., Esquerdo, G., \& Everett, M. 2011, \aj, 141, 179

\bibitem[Christensen-Dalsgaard et
al.(2010)]{Christensen-Dalsgaard:10} Christensen-Dalsgaard, J.,
Kjeldsen, H., Brown, T.\ M., et al.\ 2010, \apjl, 713, L164 

\bibitem[Christian et al.(2009)]{Christian:09} Christian,
D.\ J., Gibson, N.\ P., Simpson, E.\ K., et al.\ 2009, \mnras, 392, 1585 

\bibitem[Deleuil et al.(2008)]{Deleuil:08}
 Deleuil, M.\ et al.\ 2008, \aap, 491, 889

\bibitem[Demory \& Seager(2011)]{Demory:11}
 Demory, B.-O., \& Seager, S. 2011, \apjs, 197, 12

\bibitem[Djupvic \& Andersen(2010)]{Djupvic:10}
 Djupvic, A.\ A., \& Andersen, J. 2010, in Highlights of Spanish
 Astrophysics V, ed.\ J.\ M.\ Diego, L.\ J.\ Goicoechea, J.\ I.\
 Gonz\'alez-Serrano, \& J.\ Gorgas (Berlin: Springer), 211

\bibitem[Enoch et al.(2010)]{Enoch:10}
 Enoch, B., Collier Cameron, A., Parley, N.\ R., \& Hebb, L. 2010,
 \aap, 516, 33

\bibitem[Feigelson \& Babu(1992)]{Feigelson:92}
 Feigelson, E.\ D., \& Babu, G.\ J. 1992, \apj, 397, 55

\bibitem[Fernandez et al.(2009)]{Fernandez:09} Fernandez,
J.\ M., Holman, M.\ J., Winn, J.\ N., et al.\ 2009, \aj, 137, 4911 

\bibitem[Fortney \& Nettelmann(2010)]{Fortney:10}
 Fortney, J.\ J., \& Nettelmann, N. 2010, \ssr, 152, 423

\bibitem[F\H{u}r\'esz(2008)]{Furesz:08}
 F\H{u}r\'esz, G. 2008, PhD thesis, Univ.\ Szeged

\bibitem[Gilliland et al.(2011)]{Gilliland:11} Gilliland, R.\ L.,
McCullough, P.\ R., Nelan, E.\ P., et al.\ 2011, \apj, 726, 2

\bibitem[Gillon et al.(2010)]{Gillon:10} Gillon, M., Lanotte,
A.\ A., Barman, T., et al.\ 2010, \aap, 511, A3 

\bibitem[Girardi et al.(2000)]{Girardi:00}
 Girardi, L., Bressan, A., Bertelli, G., \& Chiosi, C. 2000, \aaps,
141, 371

\bibitem[Glazunova et al.(2008)]{Glazunova:08} Glazunova, L.\ V.,
Yushchenko, A.\ V., Tsymbal, V.\ V., Mkrtichian, D.\ E., Lee, J., J.,
Kang, Y.\ W., Valyavin, G.\ G., \& Lee, B.-C. 2008, \aj, 136, 1736

\bibitem[Hartman et al.(2011a)]{Hartman:11a} Hartman,
J.\ D., Bakos, G.\ {\'A}., Sato, B., et al.\ 2011a, \apj, 726, 52 

\bibitem[Hartman et al.(2011b)]{Hartman:11b} Hartman,
J.\ D., Bakos, G.\ {\'A}., Kipping, D.\ M., et al.\ 2011b, \apj, 728, 138 

\bibitem[Hebb et al.(2009)]{Hebb:09}
 Hebb, L.\ et al.\ 2009, \apj, 693, 1920

\bibitem[H{\'e}brard et al.(2010)]{Hebrard:10} H{\'e}brard, G.,
D{\'e}sert, J.-M., D{\'{\i}}az, R.\ F., et al.\ 2010, \aap, 516, A95

\bibitem[Hensberge et al.(2000)]{Hensberge:00} Hensberge, H.,
Pavlovski, K., \& Verschueren, W. 2000, \aap, 358, 553

\bibitem[Holman et al.(2007)]{Holman:07}
 Holman, M.\ J.\ et al. 2007, \apj, 664, 1185

\bibitem[Howard et al.(2012)]{Howard:12} Howard, J.\ A.\ et
al.\ 2012, \apj, 749, 134

\bibitem[Joshi et al.(2009)]{Joshi:09} Joshi, Y.\ C.,
Pollacco, D., Collier Cameron, A., et al.\ 2009, \mnras, 392, 1532 

\bibitem[Kipping et al.(2010)]{Kipping:10} Kipping,
D.\ M., Bakos, G.\ {\'A}., Hartman, J., et al.\ 2010, \apj, 725, 2017 

\bibitem[Kipping \& Bakos(2011a)]{Kipping:11a} Kipping, D.,
\& Bakos, G.\ 2011a, \apj, 730, 50 

\bibitem[Kipping \& Bakos(2011b)]{Kipping:11b} Kipping, D.,
\& Bakos, G.\ 2011b, \apj, 733, 36 

\bibitem[Kov\'acs et al.(2007)]{Kovacs:07}
 Kov\'acs, G.\ et al. 2007, \apj, 670, L41

\bibitem[Kov{\'a}cs et al.(2010)]{Kovacs:10} Kov{\'a}cs,
G., Bakos, G.\ {\'A}., Hartman, J.\ D., et al.\ 2010, \apj, 724, 866 

\bibitem[Kurucz(1993)]{Kurucz:93}
 Kurucz, R.\ L. 1993, ATLAS9 Stellar Atmosphere Programs and 2 km/s
 Grid CDROM 13 (Cambridge:SAO)

\bibitem[Latham et al.(2002)]{Latham:02}
 Latham, D.\ W.\ et al.\ 2002, \aj, 124, 1144

\bibitem[Latham et al.(2009)]{Latham:09} Latham, D.\ W.,
Bakos, G.\ {\'A}., Torres, G., et al.\ 2009, \apj, 704, 1107 

\bibitem[L{\'e}ger et al.(2009)]{Leger:09} L{\'e}ger, A., Rouan,
D., Schneider, J., et al.\ 2009, \aap, 506, 287 

\bibitem[McCullough et al.(2006)]{McCullough:06}
 McCullough, P.\ R.\ et al.\ 2006, \apj, 648, 1228

\bibitem[McCullough et al.(2008)]{McCullough:08}
 McCullough, P.\ R.\ et al.\ 2008, arXiv:0805.2921

\bibitem[Miller et al.(2009)]{Miller:09}
 Miller,, N., Fortney, J.\ J., \& Jackson, B. 2009, \apj, 702, 1413

\bibitem[Narita et al.(2010)]{Narita:10} Narita, N.,
Hirano, T., Sanchis-Ojeda, R., et al.\ 2010, \pasj, 62, L61 

\bibitem[Noyes et al.(2008)]{Noyes:08} Noyes, R.\ W.,
Bakos, G.\ {\'A}., Torres, G., et al.\ 2008, \apjl, 673, L79 

\bibitem[P{\'a}l et al.(2010)]{Pal:10} P{\'a}l, A., Bakos,
G.\ {\'A}., Torres, G., et al.\ 2010, \mnras, 401, 2665 

\bibitem[Perryman et al.(1997)]{Perryman:97}
 Perryman, M.\ A.\ C., et al. 1997, The Hipparcos and Tycho Catalogues
 (ESA SP-1200; Noordwjik: ESA)

\bibitem[Pollacco et al.(2006)]{Pollacco:06}
 Pollacco, D.\ L.\ et al.\ 2006, \pasp, 118, 1407

\bibitem[Press et al.(1992)]{Press:92}
 Press, W.\ H., Teukolsky, S.\ A., Vetterling, W.\ T., \& Flannery,
B.\ P. 1992, Numerical Recipes, (2nd.\ Ed.; Cambridge: Cambridge
Univ.\ Press), 650

\bibitem[Quinn et al.(2012)]{Quinn:12} Quinn, S.\ N.,
Bakos, G.\ {\'A}., Hartman, J., et al.\ 2012, \apj, 745, 80 

\bibitem[Royer et al.(2002b)]{Royer:02b} Royer, R., Grenier, S., Baylac,
M.-O., G\'omez, A.\ E., \& Zorec, J. 2002b, \aap, 393, 897

\bibitem[Royer et al.(2002a)]{Royer:02a} Royer, R., Gerbaldi, M.,
Faraggiana, R., \& G\'omez, A.\ E. 2002a, \aap, 381, 105

\bibitem[Santos et al.(2004)]{Santos:04}
 Santos, N.\ C., Israelian, G., \& Mayor, M. 2004, \aap, 415, 1153

\bibitem[Santos et al.(2006)]{Santos:06}
 Santos, N.\ C.\ et al.\ 2006, \aap, 450, 825

\bibitem[Schuler et al.(2010)]{Schuler:10}
 Schuler, S.\ C., Plunkett, A.\ L., King, J.\ R., \& Pinsonneault, M.\
 H. 2010, \pasp, 122, 766

\bibitem[Seager \& Mall\'en-Ornelas(2003)]{Seager:03}
 Seager, S., \& Mall\'en-Ornelas, G. 2003, \apj, 585, 1038

\bibitem[Shporer et al.(2009)]{Shporer:09} Shporer, A.,
Bakos, G.\ {\'A}., Bouchy, F., et al.\ 2009, \apj, 690, 1393 

\bibitem[Simpson et al.(2010)]{Simpson:10} Simpson,
E.\ K., Pollacco, D., H{\'e}brard, G., et al.\ 2010, \mnras, 405, 1867 

\bibitem[Skillen et al.(2009)]{Skillen:09} Skillen, I.,
Pollacco, D., Collier Cameron, A., et al.\ 2009, \aap, 502, 391 

\bibitem[Smalley et al.(2001)]{Smalley:01}
 Smalley, B., Smith, K., \& Dworetsky, M. 2001, \hfill\linebreak
{\tt http://www.astro.keele.ac.uk/\textasciitilde bs/publs/uclsyn.pdf}

\bibitem[Sneden(1973)]{Sneden:73}
 Sneden, C.\ A. 1973, Ph.D.\ thesis, Univ.\ Texas (Austin)

\bibitem[Southworth(2009a)]{Southworth:09a} Southworth, J.\ 2009a,
\mnras, 394, 272 

\bibitem[Southworth et al.(2009b)]{Southworth:09b} Southworth,
J., Hinse, T.\ C., Dominik, M., et al.\ 2009b, \apj, 707, 167 

\bibitem[Souza et al.(2007)]{Sousa:07}
 Sousa, S.\ G., Santos, N.\ C., Israelian, G., Mayor, M., \& Monteiro,
 M.\ J.\ P.\ F.\ G. 2007, \aap, 469, ,783

\bibitem[Sozzetti et al.(2006)]{Sozzetti:06}
 Sozzetti, A., Yong, D., Carney, B.\ W., Laird, J.\ B., Latham, D.\
 W., \& Torres, G. 2006, \aj, 131, 2274

\bibitem[Sozzetti et al.(2007)]{Sozzetti:07}
 Sozzetti, A., Yong, D., Torres, G., Charbonneau, D., Latham, D.\ W.,
 Holman, M.\ J., Winn, J.\ N., Laird, J.\ B., \& O'Donovan, F.\ T.
 2007, \apj, 664, 1190

\bibitem[Sozzetti et al.(2009)]{Sozzetti:09}
 Sozzetti, A.\ et al.\ 2009, \apj, 691, 1145

\bibitem[Street et al.(2010)]{Street:10} Street, R.\ A.,
Simpson, E., Barros, S.\ C.\ C., et al.\ 2010, \apj, 720, 337 

\bibitem[Torres et al.(2002)]{Torres:02}
 Torres, G., Neuh\"auser, R., \& Guenther, E.\ W. 2002, \aj, 123, 1701

\bibitem[Torres et al.(2007)]{Torres:07} Torres, G., Bakos,
G.\ {\'A}., Kov{\'a}cs, G., et al.\ 2007, \apjl, 666, L121 

\bibitem[Torres et al.(2008)]{Torres:08}
 Torres, G., Winn, J.\ N., \& Holman, M.\ J. 2008, \apj, 677, 1324

\bibitem[Torres et al.(2010b)]{Torres:10b} Torres, G.,
Bakos, G.\ {\'A}., Hartman, J., et al.\ 2010b, \apj, 715, 458 

\bibitem[Torres et al.(2010a)]{Torres:10a}
 Torres, G., Andersen, A., \& Gim\'enez, A. 2010a, A\&ARv, 18, 67

\bibitem[Valenti \& Fischer(2005)]{Valenti:05}
 Valenti, J.\ A., \& Fischer, D.\ A. 2005, \apjs, 159, 142

\bibitem[Valenti \& Piskunov(1996)]{Valenti:96}
 Valenti, J.\ A., \& Piskunov, N. 1996, \aaps, 118, 595

\bibitem[van Leeuwen(2007)]{vanLeeuwen:07}
 van Leeuwen, F. 2007, Hipparcos, the New Reduction of the Raw Data,
Astr.\ and Sp.\ Sci.\ Library, Vol.\ 350 (Cambridge: Springer)

\bibitem[Vogt et al.(1994)]{Vogt:94}
 Vogt, S.\ S.\ et al. 1994, Proc.\ SPIE, 2198, 362

\bibitem[West et al.(2009)]{West:09} West, R.\ G.,
Collier Cameron, A., Hebb, L., et al.\ 2009, \aap, 502, 395 

\bibitem[Winn et al.(2008)]{Winn:08} Winn, J.\ N.,
Holman, M.\ J., Torres, G., et al.\ 2008, \apj, 683, 1076 

\bibitem[Yi et al.(2001)]{Yi:01}
 Yi, S.\ K., Demarque, P., Kim, Y.-C., Lee, Y.-W., Ree, C.\ H.,
Lejeune, T., \& Barnes, S. 2001, \apjs, 136, 417

\bibitem[Yong et al.(2004)]{Yong:04}
 Yong, D., Lambert, D.\ L., Allende Prieto, C., \& Paulson, D.\
 B. 2004, \apj, 603, 697

\end{thebibliography}
\end{document}